\documentclass[letter,12pt]{article}
\pdfoutput=1
\usepackage[table]{xcolor}
\usepackage{relsize}
\usepackage{bm}
\usepackage{algorithm}
\usepackage{algpseudocode}
\usepackage{mathrsfs}

\usepackage[left=1in, top=1in, bottom=1in, right=1in]{geometry}

\usepackage{latexsym,amssymb,amsmath,color}
\usepackage{graphicx}
\usepackage{cite}
\usepackage[table]{xcolor}

\newcommand{\be}{\begin{equation}}
\newcommand{\ee}{\end{equation}}
\newcommand{\bea}{\begin{eqnarray}}
\newcommand{\eea}{\end{eqnarray}}

\def\1{{\rm 1}}

\def\s1{^{\rm (1)}}

\def\1{{\rm 1}}

\def\s1{^{\rm (1)}}

 \newcommand{\argmax}{\operatornamewithlimits{arg\,max}}
 
\usepackage[symbol*]{footmisc}
\DefineFNsymbolsTM{myfnsymbols}{
  \textdagger    \dagger
  \textdaggerdbl \ddagger
  \textsection   \mathsection
  \textbardbl    \|%
  \textparagraph \mathparagraph
}%
\setfnsymbol{myfnsymbols}

\usepackage[latin1]{inputenc}
\usepackage{tikz}
\usetikzlibrary{shapes,arrows}
    \tikzstyle{blockr} = [rectangle, draw, fill=blue!20,
    text width=9.0em, text centered, rounded corners]
    \tikzstyle{block} = [rectangle, draw=none, fill=blue!20, anchor=north, 
    text width=9.0em, text centered]
\tikzstyle{line} = [draw, -latex']

\begin{document}



\begin{center}
\begin{Large}
{\bf Design Analysis for Optimal Calibration of Diffusivity in Reactive Multilayers} \\
\end{Large}

\bigskip

Manav Vohra,$^1$ Xun Huan,$^2$ Timothy P. Weihs,$^{3}$ Omar M. Knio$^{1,4,}\footnote{Corresponding Author (Email: omar.knio@duke.edu)}$ \\

\bigskip
\bigskip

\normalsize
$^1$Department of Mechanical Engineering and Materials Science\\
Duke University\\
Durham, NC 27708\\

\bigskip

$^2$Department of Aeronautics and Astronautics\\
Massachusetts Institute of Technology\\
Cambridge, MA 02139\\

\bigskip

$^3$Department of Materials Science and Engineering\\
The Johns Hopkins University\\
Baltimore, MD 21218\\

\bigskip

$^4$Computer, Electrical and Mathematical Sciences and Engineering Division \\
King Abdullah University of Science and Technology \\
Thuwal 23955-6900, Kingdom of Saudi Arabia\\

\end{center}


%

%
%
%


\section*{Abstract}

Calibration of the uncertain Arrhenius diffusion parameters for
quantifying mixing rates in Zr-Al nanolaminate foils was performed in
a Bayesian setting~\cite{Vohra:2014b}. The parameters were inferred in
a low temperature regime characterized by homogeneous ignition and a
high temperature regime characterized by self-propagating reactions in
the multilayers. In this work, we extend the analysis to find optimal
experimental designs that would provide the best data for
inference. We employ a rigorous framework that quantifies the expected
information gain in an experiment, and find the optimal design
conditions using numerical techniques of Monte Carlo, sparse
quadrature, and polynomial chaos surrogates.  For the low temperature
regime, we find the optimal foil heating rate and pulse duration, and
confirm through simulation that the optimal design indeed leads to
sharper posterior distributions of the diffusion parameters. For the
high temperature regime, we demonstrate potential for increase in the
expected information gain of the posteriors by increasing sample size
and reducing uncertainty in measurements. Moreover, posterior
marginals are also produced to verify favorable experimental scenarios
for this regime.

\clearpage


\section{Introduction}
\label{sec:intro}

Sputter-deposited reactive multilayers are characterized by bilayers
of approximately uniform thickness that alternate between distinct
elements and mix exothermically.  Typically, the bilayer thickness is
on the order of tens of nanometers, whereas the overall multilayer
thickness is tens of microns.  The multilayers support
self-propagating reactions that travel along the direction of the
layers at speeds ranging from several centimeters per second to tens
of meters per second~\cite{Blobaum:2003, Kim:2011, Trenkle:2005,
  Trenkle:2008}.  These special properties make them attractive for a
wide range of applications, including joining processes such as
welding, soldering and brazing~\cite{Swiston:2005,Duckham:2008,
  Vanheerden:2008, Wang:2008,Vanheerden:2006a,
  Vanheerden:2006b,Merzhanov:1990,Koizumi:1990}, and ignition of
secondary reactions and defeat of biocidal
agents~\cite{Joress:2012, Vohra:2014, Overdeep:2015, Feng:2013,
  Sullivan:2010, Gavens:2000, Besnoin:2002, Ma:1990}. Advances in
fabrication techniques~\cite{Blobaum:2003, Gavens:2000, Reiss:1999,
  Wang:2004, Rogachev:2004, Gachon:2005, Adams:2006, Adams:2006a}
enabling control of the microstructure of metallic multilayers at the
atomistic levels, as well as simultaneous discovery of novel
applications, have motivated a sustained development of reaction
models for more than three decades.  In particular, model calibration
has typically relied on data from experiments conducted on
geometrically flat nanolaminate foils comprising hundreds of bilayers.
Once validated, the model predictions are then exploited to
characterize reaction properties, and to identify optimal
microstructure, composition, and geometry suited for targeted
application scenarios.

Early reaction models often used a conserved scalar, $c$, to describe the evolution of formation reactions in metallic 
multilayers~\cite{Jayaraman:1998a, Mann:1997, Munir:1990}.  The conserved scalar, which measures the 
degree of mixing, is assumed to evolve according to the Fickian diffusion law:
\be
\frac{\partial c}{\partial t} = \nabla \cdot (D\nabla c)
\label{eq:fick}
\ee
where $D$ is the atomic diffusivity.  The reaction kinetics strongly depend on the diffusivity and its variation with temperature.
The latter is typically represented in terms of the following Arrhenius law:
\be
D = D_{0}\exp\left(-\frac{E_{a}}{RT}\right)
\label{eq:diff}
\ee
where $D_{0}$ is the pre-exponent and $E_{a}$ is the activation energy.
This original formulation from~\cite{Jayaraman:1998a} was extended to
incorporate melting effects in reactants and products during the
progress of reaction~\cite{Besnoin:2002}.  In order to overcome
limitations due to the stiffness of the governing equations, Salloum
and Knio recently proposed a reduced reaction formalism, and
demonstrated orders of magnitude enhancements in computational
efficiency for transient multidimensional simulations of Ni-Al
multilayers~\cite{Salloum:2010a, Salloum:2010b, Salloum:2010c}.  The
reduced model was further generalized to account for variable thermal
transport properties, and used to examine the behavior of
self-propagating reaction fronts~\cite{Alawieh:2011, Sraj:2013}.


Recently, Fritz~\cite{Fritz:2011} studied homogenous reactions in
Ni-Al nanolaminate foils that are triggered using a uniform current
pulse.  It was observed in these experiments that the measured
intermixing rates were significantly smaller than those predicted by
global Arrhenius atomic diffusivity law calibrated based on
self-propagating velocity measurements.  This observed discrepancy
implies that a single global Arrhenius diffusivity law over the entire
range of reaction temperatures is inadequate. Consequently,
Alawieh~\emph{et al.}~\cite{Alawieh:2013} introduced a composite
diffusivity law that comprises two Arrhenius branches, describing
the evolution of diffusivity at temperature regimes below and above
the melting point of Al.  The low-temperature branch was calibrated
based on ignition experiments, whereas the high-temperature branch was
inferred from measurements of self-propagating front velocities.  For
both branches, least-squares regression techniques were used to
estimate the unknown Arrhenius parameters.  Using the new composite
diffusivity law,~\cite{Alawieh:2013} demonstrated the possibility of
using a single model to capture various experimental phenomena,
including homogeneous ignition, self-propagation, and nanocalorimetry.

Vohra~\emph{et al.}~\cite{Vohra:2014b} further generalized the
methodology developed in~\cite{Alawieh:2013} by adapting a Bayesian
framework for inferring the Arrhenius parameters.  The calibration
step was accelerated by constructing functional representations of
experimental observables, judiciously selected so that they exhibit
smooth dependence on the uncertain model parameters.  In particular,
polynomial surrogates were constructed for reaction temperature in
ignition experiments, and for reaction velocity in self-propagating
front experiments.  The overall method was applied to calibrate a
composite Arrhenius diffusivity law for Zr-Al multilayers.  Experience
from the study in~\cite{Vohra:2014b} also highlighted the potential
impact of data noise and scarcity on residual uncertainties in
inferred quantities.


In this work, we aim to demonstrate significant potential for application
of an optimal Bayesian
experimental design framework~\cite{Huan:2013} to guide the selection of
test conditions (experimental scenarios) in intermixing calibration
 experiments for reactive multilayers.
Consequently, the experimental data is expected to significantly
enhance the information gain in mixing rates and hence offer a robust
strategy for model calibration.  
Specifically, we focus our attention on low-temperature homogeneous ignition
experiments and self-propagating front experiments.  Description of
the reaction models used for both setups can be found in
Section~\ref{sec:method}.
Section~\ref{sec:framework} introduces the optimal Bayesian
experimental design framework, where an objective function (expected
utility) is formulated to quantify and reflect the average information
that can be gained from an experiment, on the uncertain model
parameters. Furthermore, this quantity is estimated using Monte Carlo
sampling, and requires a very high number of reaction model
evaluations, making it impractical.  To mitigate the costs,
computationally affordable polynomial chaos (PC) surrogates are
utilized to accelerate both the Bayesian design and inference phases.

As mentioned earlier, the experimental design framework is applied to
two Zr-Al reaction
regimes as discussed in much detail in Section~\ref{sec:results}.  First, a low temperature
regime where a uniform current pulse of finite width is used to
trigger homogeneous ignition of the nanolaminate.  For this
application, we use multilayers with a fixed bilayer width,
and the design variables are the amplitude and duration of the current
pulse. In additional to finding the optimal experimental design, we
also perform Bayesian inference at optimal and non-optimal conditions,
and demonstrate the higher information gain in the former.
Second, a self-propagating reaction front regime, where the
reaction front propagates parallel to the bilayers.
In this case, the premix-width and the initial temperature of the Zr-Al
nanolaminate foil are considered as design variables. Average information
gain in the Arrhenius diffusion parameters for this regime is analyzed
for a broad range of design conditions, sample size, and uncertainty in
the data. Results based on varying either the bilayer thickness, the premix-width,
or the initial Zr-Al foil temperature are compared and contrasted in Section~\ref{ssec:prop}.
Furthermore, our findings from Bayesian inference based on synthetic data
reflecting repeated measurements at two extreme values of the bilayer thickness
 and synthetic
data collected at regular intervals throughout the considered range of bilayer thicknesses
are compared.
A posterior predictive check is also
performed to verify the suitability of the predictions in this regime.
Finally, conclusions are presented in Section~\ref{sec:conc}.


\section{Reaction Models}
\label{sec:method}

We rely on the reduced formalism proposed by Salloum and
Knio~\cite{Salloum:2010a, Salloum:2010b, Salloum:2010c} to model the
dynamics of the reaction.  In the reduced model, the canonical
solution of the diffusion equation is used to replace the partial
differential equation (PDE)~\eqref{eq:fick}, resulting in an ordinary
differential equation (ODE) for the ``normalized" age, $\tau$, of the
mixed layer: \be \frac{d\tau}{dt} = \frac{D(T)}{\delta^{2}}
\label{eq:tau}
\ee where $\delta$ is thickness of the Al layer, and $D(T)$ is the
diffusivity as a function of temperature.  \eqref{eq:tau} is further
accompanied by energy balance, which is expressed in terms of an
enthalpy evolution equation that accounts for chemical heat release,
thermal transport, and heat transfer.  The resulting coupled ODE
system is then integrated in time to simulate the evolution of the
physics.  The formulation of the enthalpy equation is briefly
described in Section~\ref{sub:ignition} for the homogeneous reactions,
and in Section~\ref{sub:propagation} for self-propagating reactions.

\subsection{Homogeneous Ignition}
\label{sub:ignition}

For homogeneous ignition, the enthalpy and temperature of
the multilayer can be treated as spatially uniform.  For reactions
initiated with a current pulse, we may use a simplified energy balance
equation for thin multilayer foils:
\be
\frac{d H}{dt} =
\frac{V\sp{\prime}I}{V_{foil}}\mathbf{H}[t_{p}-t] + \frac{\partial
  \overline{Q}(\tau)}{\partial t} - \frac{h}{d}(T - T_{0})
\label{eqn:enthalpy1}
\ee where $H$ is the section-averaged enthalpy, $V\sp{\prime}$ and $I$
are the respective voltage and current applied across the foil, $t$ is
time, $t_p$ is the width of the uniform current pulse, $\mathbf{H}$ is
the Heaviside function, $\overline{Q}$ is chemical heat release, $d$
is the foil thickness, $h$ is the heat transfer coefficient, and $T$
and $T_0$ denote the instantaneous foil temperature and ambient
temperature, respectively.  The temperature is recovered from the
enthalpy by inverting a temperature-enthalpy relationship that
accounts for phase transformations, and accordingly involves the heats
of fusion of the reactants and
products~\cite{Besnoin:2002,Vohra:2014b}.
 
\subsection{Self-propagating Reactions}
\label{sub:propagation}

For self-propagating reactions, our formulation of the energy
conservation equation assumes that the reaction front propagates parallel to
 the deposited bilayers~\cite{Blobaum:2003, Kim:2011,
  Trenkle:2005, Trenkle:2008}, and that heat losses are negligible
during self-propagation.  These assumptions are suitable for most
experimental setups for measuring the front velocity, which typically
rely on igniting self-propagating fronts in thick foils with fine,
nanoscale layering~\cite{Jayaraman:2001}.

Under the above assumptions, the energy balance equation can be
expressed as: \be \frac{\partial H}{\partial t} =
\overline{k}\frac{\partial^{2}T}{\partial x^{2}} + \frac{\partial
  \overline{Q}(\tau)}{\partial t} \ee where $H$ is the
section-averaged enthalpy, $x$ is the direction of propagation, and
$\overline{k}$ is the molar averaged
thermal conductivity of the constituents.

For both types of reaction, we use the reduced model
formalism~\cite{Salloum:2010a} to estimate the chemical heat release
term.  For self-propagating fronts, the formulation incorporates
temperature-dependent correlations for the specific heat and thermal
conductivity.  Additional details concerning thermo-physical
properties and the numerical integration methodologies are provided
in~\cite{Vohra:2014b}.


\section{Optimal Bayesian Experimental Design}
\label{sec:framework}

{Optimal} experimental design involves using a model to help find the
best choice of designs for a targeted experimental purpose or goal. We
are particularly interested in performing design for nonlinear models,
where the model observables (or quantities of interest, QoIs) depend
nonlinearly on the uncertain parameters. For brevity, we do not
provide a full introduction of this rich field here, but instead refer
interested readers to~\cite{Chaloner:1995, Muller:2005}, as well as
literature review within~\cite{Huan:2013, Huan:2014}.

We adopt the methodology developed in~\cite{Huan:2013} to find optimal
design conditions for our experiments. We take a Bayesian perspective
(e.g.,~\cite{Berger:1985, Epstein:1985, Sivia:1996}), and characterize
the uncertainty in variables using probability density functions
(PDFs)---i.e., treat them as random variables. Bayes' theorem thus
provides a natural mechanism for ``updating'' our knowledge about the
uncertainty of parameters $\bm{X}$ when new data $\bm{y}$ are obtained
from performing an experiment at a design condition $\bm{d}$:
\begin{eqnarray}
  p(\bm{X} | \bm{y}, \bm{d}) = \frac{p(\bm{y} | \bm{X},
    \bm{d})p(\bm{X})}{p(\bm{y}|\bm{d})}\label{e:Bayes_original}
\end{eqnarray}
where $p(\cdot)$ denotes PDF, $p(\bm{X})$ is the prior, $p(\bm{y} |
\bm{X}, \bm{d})$ is the likelihood, $p(\bm{X} | \bm{y}, \bm{d})$ is
the posterior, and $p(\bm{y}|\bm{d})$ is the evidence term that
normalizes the posterior PDF.  Intuitively, then, for example, we
would like to perform an experiment whose observations result in
``tight'' or ``sharp'' posterior distributions, i.e., high information
gain and uncertainty reduction.

To mathematically quantify the usefulness or value of an experiment,
we introduce in Section~\ref{sub:utility} an objective function (also
called the expected utility function in the experimental design
context) that is to be maximized. This quantity can only be estimated
through numerical methods, which typically involve a high number of
model evaluations. In order to overcome this extremely computationally
expensive (if not altogether intractable) requirement, we employ
polynomial chaos (PC) expansions to act as surrogate models in place of the
original ODE system. While they are approximations, PC surrogates can
efficiently capture the dependence of QoIs on both the design variables and
uncertain parameters, and greatly accelerate the computations in the
design optimization as well as the post-experiment Bayesian
inference/calibration phases. A brief introduction to the PC methodology
is presented in Section~\ref{sub:surrogate}.

\subsection{Expected Utility}
\label{sub:utility}

There are different possible approaches for quantifying the usefulness
of data gathered from an experiment. For instance, one may take a
goal-oriented approach, and impose a loss function based on specific
desirable mission objectives from the experiments. More broadly, when
the precise experimental goals are unknown, or if one wishes to design
experiments that are generally good for a range of possible
objectives, one may choose to simply learn about (i.e., reduce) the
uncertainty associated with the physical system. We take this
approach, and devise an objective function reflecting the
\textit{expected information gain} on the uncertain parameters from an
experiment, based on the Kullback-Leibler (KL) divergence from
posterior to prior~\cite{Lindley:1956, Lindley:1972}:
\begin{equation}
  U({\bm d}) =
  \mathbb{E}_{\bm{y}|\bm{d}}\left[D_{\textnormal{KL}}(p(\bm{X}|\bm{y},\bm{d})
    \|p(\bm{X}))\right].
  \label{eqn:utility}
\end{equation}
Since the observations are unknown when we design the experiments, we
thus take an expectation $\mathbb{E}_{\bm{y}|\bm{d}}$ with respect to
the data $\bm{y}$ given the particular candidate design $\bm{d}$.
 The final $U(\bm{d})$ is thus called the
\textit{expected utility} of performing an experiment at design
condition $\bm{d}$. The optimal design problem thus involves finding
\begin{eqnarray}
  \bm{d}^{\ast} = \argmax U({\bm d}).
\end{eqnarray}

The expected utility in \eqref{eqn:utility} generally has no
analytical form for models with nonlinear parameter-output
relationships, and one can only approximate it numerically. We use a
doubly-nested Monte Carlo (MC) estimator~\cite{Ryan:2003, Huan:2013}:
\begin{equation}
  U(\bm{d}) \approx \frac{1}{N_{out}} \sum_{i=1}^{N_{out}} \left\{
  \ln[p(\bm{y}^{(i)}|\bm{X}^{(i)},\bm{d})] -
  \ln[p(\bm{y}^{(i)}|\bm{d})]\right\}
  \label{eqn:outer}
\end{equation}
where $N_{out}$ denotes the number of samples used for this ``outer''
loop, $\bm{X}^{(i)}$ are parameter samples drawn from the prior, and
$\bm{y}^{(i)}$ are observation samples generated from the likelihood
given $\bm{X}^{(i)}$ and $\bm{d}$. An ``inner'' MC estimator is also
used for estimating the evidence (second) term
$p(\bm{y}^{(i)}|\bm{d})$ in~\eqref{eqn:outer}:
\begin{equation}
  p(\bm{y}^{(i)}|\bm{d}) \approx \frac{1}{N_{in}} \sum_{j=1}^{N_{in}}
  p(\bm{y}^{(i)}|\bm{X}^{(i,j)},\bm{d})
  \label{eqn:inner} 
\end{equation}
where $N_{in}$ denotes the number of inner MC samples, and
$\bm{X}^{(i,j)}$ are samples from the prior.  A direct application of
the doubly-nested MC estimators in~\eqref{eqn:outer}
and~\eqref{eqn:inner} together has complexity
$\mathcal{O}(N_{out}N_{in})$. We follow the sampling technique
introduced in~\cite{Huan:2013}, and reuse samples across the two MC
loops to reduce the computations to $\mathcal{O}(N_{out})$.

Finally, we note that the estimators in~\eqref{eqn:outer}
and~\eqref{eqn:inner} only require evaluations of the likelihood
$p(\bm{y}|\bm{X},\bm{d})$. For this, we assume an additive Gaussian
noise model:
\begin{equation}
  \bm{y} = \bm{M}(\bm{X},\bm{d}) + \bm{\epsilon}
  \label{eqn:disc}
\end{equation}
where $\bm{M}(\bm{X},\bm{d})$ is the model (i.e., ODE system)
evaluation at the specified $\bm{X}$ and $\bm{d}$. Gaussian noise, 
$\bm{\epsilon}\sim \mathcal{N}(\bm{0},\bm{\Sigma})$ with the covariance matrix
$\bm{\Sigma}$ assumed to be diagonal (i.e.,
componentwise-independent noise) with components $\sigma_{i}$ being a
fixed fraction of the corresponding model output (i.e., $\sigma_{i}$ =
$\alpha M_{i}$, $\alpha$ scalar constant, so $\alpha=0.1$ means a 10\%
noise). Estimating the expected utility at a given design point hence 
involves $N_{out}$ model evaluations which can be computationally
intensive especially for a large number of design considerations. 
As discussed earlier, we rely on polynomial chaos surrogates to
estimate model evaluations in order to mitigate computational
costs. 

\subsection{Polynomial Chaos (PC) Surrogate}
\label{sub:surrogate}

PC is a spectral expansion technique that uses polynomials to capture the
statistical dependence of model outputs on inputs, and are
computationally much cheaper to evaluate.  The PC methodology
parameterizes random variables in terms of canonical random variables
$\bm{\xi} \in \Omega$, and expresses the dependence of a generic QoI,
$Q$, on $\bm{\xi}$, in terms of a mean-square convergent expansion of
the form~\cite{Wiener:1938, Ghanem:1991, Olivier:2010}: \be
Q(\bm{\xi}) = \sum_{\bm{k}=0}^{P} c_{\bm{k}}\Psi_{\bm k}(\bm{\xi})
\label{eq:QoI}
\ee where $P$ is the number of terms retained in the truncated basis,
the $\Psi_{\bm k}$'s form an orthogonal basis of the Hilbert space
with the property that a member function, $f(\bm \xi)$, is
bounded:
\begin{eqnarray}
  L_2(\Omega) = \left\{ f: \Omega \to {\mathbb R} \left| \int_\Omega
  f^2(\bm{\xi}) d\mu(\bm{\xi}) < \infty \right.\right\}
\end{eqnarray}

For the application in this study, the uncertain parameters and design
variables are endowed with uniform distributions (details in
Section~\ref{sec:results}).  Accordingly, the components of $\bm{\xi}$
are independent, identically distributed uniform random variables over
$[-1,1]$, and the $\Psi_{\bm k}$'s are multi-dimensional Legendre
polynomials defined over $\Omega = [-1,1]^d$ with $d$ being the
dimension of $\bm{\xi}$ reflecting the system's stochastic degrees of
freedom~\cite{Xiu:2002, Olivier:2010}.  Note that we also rely on the
PC expansion to represent dependence of QoIs on the design variables
$\bm d$.  In other words, the representation in (\ref{eq:QoI}) is a
function of \textit{both}, $\bm X$ and $\bm d$~\cite{Huan:2013}.

A deterministic approach is used to compute the unknown coefficients,
$c_{\bm{k}}$, in the PC representation.  The adaptive pseudo-spectral
projection (aPSP) technique~\cite{Constantine:2012,Conrad:2013,
  Winokur:2013} is used for this purpose.  The aPSP method relies on
nested sparse grids based on a Smolyak
tensorization~\cite{Smolyak:1963}.  The algorithm exploits the use of
refinement indicators to selectively enrich the resolution along
uncertain dimensions.  Meanwhile, the pseudo-spectral projection
technique ensures that the representation retains the largest number
of polynomials whose coefficients can be determined without internal
aliasing on the anisotropic sparse grid.


Once the PC surrogate is constructed, it not only accelerates the
optimal Bayesian experimental design procedure, but also contributes
to the overall uncertainty quantification analysis through global
sensitivity analysis (GSA) and post-experiment Bayesian inference.
GSA is a framework that assesses the sensitivity of QoIs with respect
the uncertain parameters and design variables, across their
\textit{entire} space (hence ``global''). One popular approach
involves the computation of Sobol sensitivity
indices~\cite{Crestaux:2009,
  Homma:1996,Olivier:2010,Sobol:2001,Sudret:2008,Alexanderian:2012},
which quantify the input variable variance contributions toward the
QoI variance. Sobol indices enable us to assess the relative
importance of uncertain parameters and design variables, and to gain
valuable insight into the design problem even before conducting more
elaborate and computationally demanding inverse problems. The Sobol
indices can be analytically extracted directly from the PC
coefficients as discussed further in Section~\ref{sec:results}.

Once the experimental data or observations are made
available, PC surrogates can also be used to accelerate the
Bayesian inference of the uncertain model parameters. Specifically, we
are interested in computing the so-called posterior distribution
of $\bm X$ using observations, $\bm y$ corresponding to design
conditions, $\bm{d}^{\ast}$ using the Bayes' rule:

\be p(\bm {X},\sigma^2 | \bm{y},\bm{d}^{\ast}) = \frac{p(\bm{y}|\bm
  {X},\sigma^2,\bm{d}^{\ast})p(\bm{X})
  p(\sigma^2)}{p(\bm{y}|\bm{d}^{\ast})}, \ee
where $p(\sigma^2)$ is the prior distribution of the hyper-parameter.
An uninformative, Jeffrey's prior~\cite{Sivia:1996} for $\sigma^2$ is
assumed: $p(\sigma^{2}) = 1/\sigma^{2}$.  The \textit{joint} posterior
distribution is estimated using Markov chain Monte Carlo (MCMC)
technique---specifically an adaptive Metropolis
algorithm~\cite{Haario:2001}---which involves repeated sampling of the
likelihood.  As outlined in~\cite{Marzouk:2007,Marzouk:2009}, using
the surrogate in estimating the likelihood yields significant
advantages, as the cost of performing a forward model run is replaced
with that of evaluating a PC representation.


\section{Results}
\label{sec:results}

We now apply the optimal Bayesian experimental design framework to
find the most informative design conditions for homogeneous ignition
and self-propagation experiments.

\subsection{Homogeneous Ignition}
\label{ssec:ignition}

We consider a nanolaminate with bilayer thickness, $\lambda=68$~nm,
premix width, $w_{0} = 0.8$~nm, and design an experiment aimed at
producing the largest expected information gain on the pre-exponent,
$D_0$, and the activation energy, $E_a$, by selecting the pulse duration,
$t_p$, and heating rate, $dT/dt$.  The parameters are assumed to have
uniform priors in the following ranges: $D_{0} \in [0.6\times
  10^{-10}, 1.20\times 10^{-10}]$~m$^{2}$/s and $E_{a} \in [47,
  53]$~kJ/mol. The design variables are similarly assigned uniform
measures\footnote{While design variables are physically not random,
  their assigned distributions serve as a weight measure of PC
  accuracy in the design space. See~\cite{Huan:2013} for details.}:
$dT/dt \in [2\times 10^{4}, 3\times 10^{4}]$~K/s and $t_{p} \in [8,
  12]$~ms. Our QoI is the reaction time, $t_r$, defined as the time
elapsed from the onset of the current pulse until the foil reaches a
prescribed temperature value $T_r$. Unless otherwise stated, this
threshold is assumed to be 900~K.
Table~\ref{t:all_variables} summarizes all the variables.

\begin{table}[htb]
\begin{center}
\begin{tabular}{c|cc}
\hline
 &  Homogeneous Ignition  & Self-Propagation \\
\hline
Uncertain parameters, $\bm{X}$ & $D_0$, $E_a$ & $D_0$, $E_a$ \\
Design variables, $\bm{d}$ & $t_p$, $dT/dt$ & $T_0$, $w_0$\\
Observable/QoI, $\bm{y}$ & $t_r$ & $v$\\
\hline 
\end{tabular}
\end{center}
\caption{Summary of uncertain parameters $\bm{X}$, design variables
  $\bm{d}$, and observable/QoI $\bm{y}$ of the homogeneous ignition
  and self-propagation experiments.}
\label{t:all_variables}
\end{table}

A PC surrogate is constructed for the QoI as a function of the total
four degrees of freedom (two from $\bm{X}$ and two from $\bm{d}$)
represented by uniform canonical variables $\bm{\xi}=[\xi_1, \xi_2,
  \xi_3, \xi_4]$ and Legendre polynomial basis:
\be t_r = \sum_{k=0}^P c_k\Psi_k(\bm{\xi}). \ee
The coefficients $c_k$
are computed using aPSP based on a nested Gauss-Patterson-Kronrod
(GPK) quadrature rule~\cite{Patterson:1968}, which required 40
sparse-grid refinement steps and a total of 1633 model simulations
performed at nodes of the adapted sparse grid. The four-dimensional
polynomial basis of the resulting surrogate was found to be of the
order: 3, 2, 2 and 2 along $\xi_1$, $\xi_2$, $\xi_3$ and $\xi_4$ respectively.

The reaction time $t_r$
has been observed to increases monotonically toward $T_r$ when $T_r$
falls in the range of temperatures extending from ignition threshold
to the melting point of Al, and $t_r$ also exhibits a smooth
dependence on diffusivity parameters~\cite{Vohra:2014b}. We thus
expect the PC expansion of $t_r$ to converge rapidly, and a low-order
expansion would be sufficient. To verify the quality of PC surrogate,
we compute the following ``worst-case'' relative error:
\be \mbox{Relative Error} =\max_{1\leq m \leq N_r} \frac{ |Q_{m} -
  Q_{s}|}{|Q_{m}|}
\label{eqn:relerror}
\ee
where $Q_{m}$, $m=1,\ldots,N_r$, is the QoI evaluated from the ODE
model at the $m$th quadrature point, and $Q_s$ denotes the
corresponding values obtained from the PC surrogate.  The relative
error of our PC surrogate is 3.16$\times$10$^{-5}$, confirming its
suitable accuracy.  The same adapted sparse grid is then used to
construct PC surrogates for $t_r$ with $T_r = 600$~K, 700~K, and
800~K.  The relative errors for these surrogates are
1.41$\times$10$^{-4}$, 1.12$\times$10$^{-4}$ and 4.47$\times$10$^{-5}$
respectively, indicating that the adapted sparse grid also leads to
suitable representations of reaction times across a broad range of
reaction temperature thresholds.

\begin{figure}[htbp]
 \begin{center}
 \begin{tabular}{cc}
\includegraphics[width=0.5\textwidth]{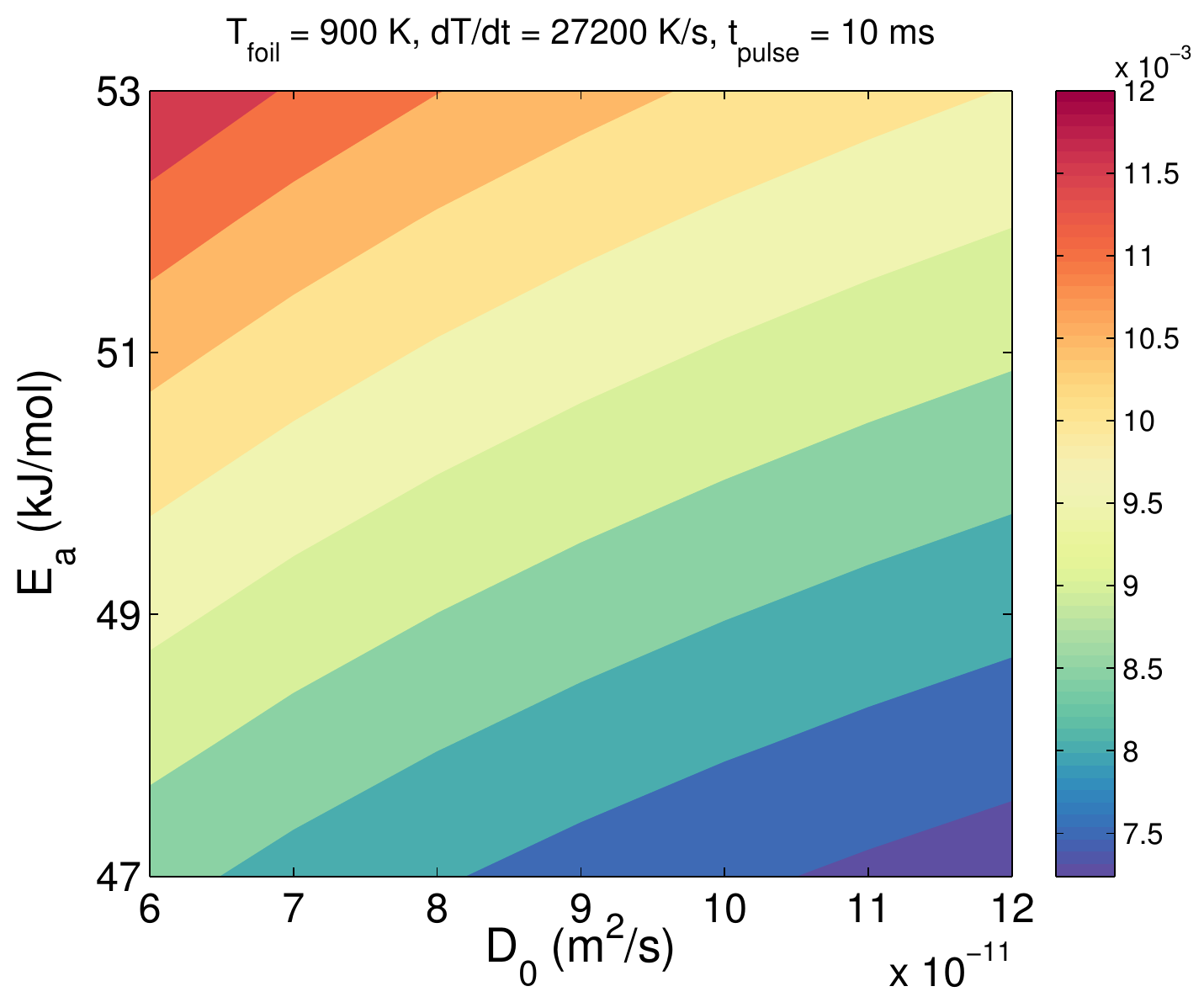}
&
\includegraphics[width=0.5\textwidth]{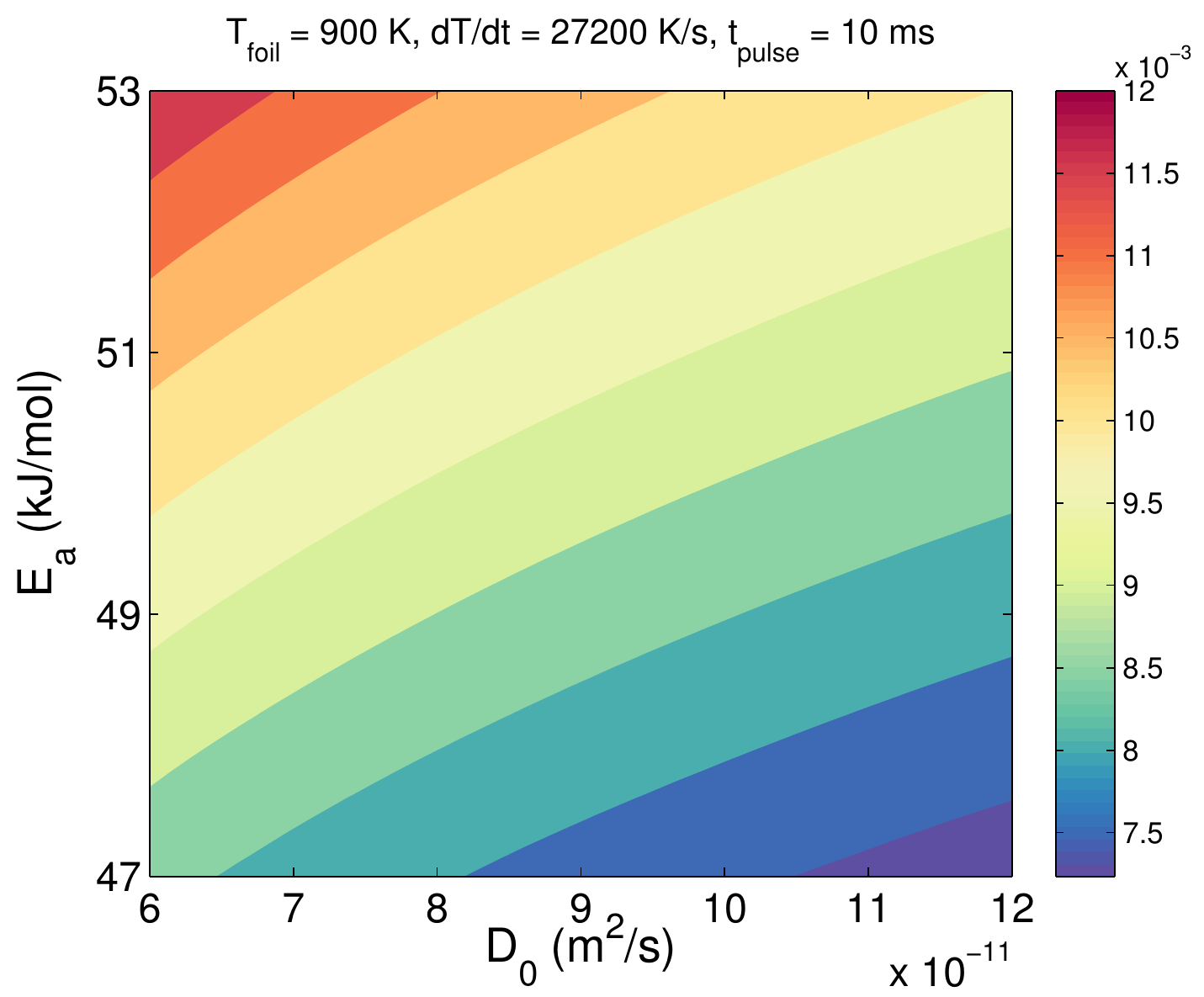}
\\
\includegraphics[width=0.5\textwidth]{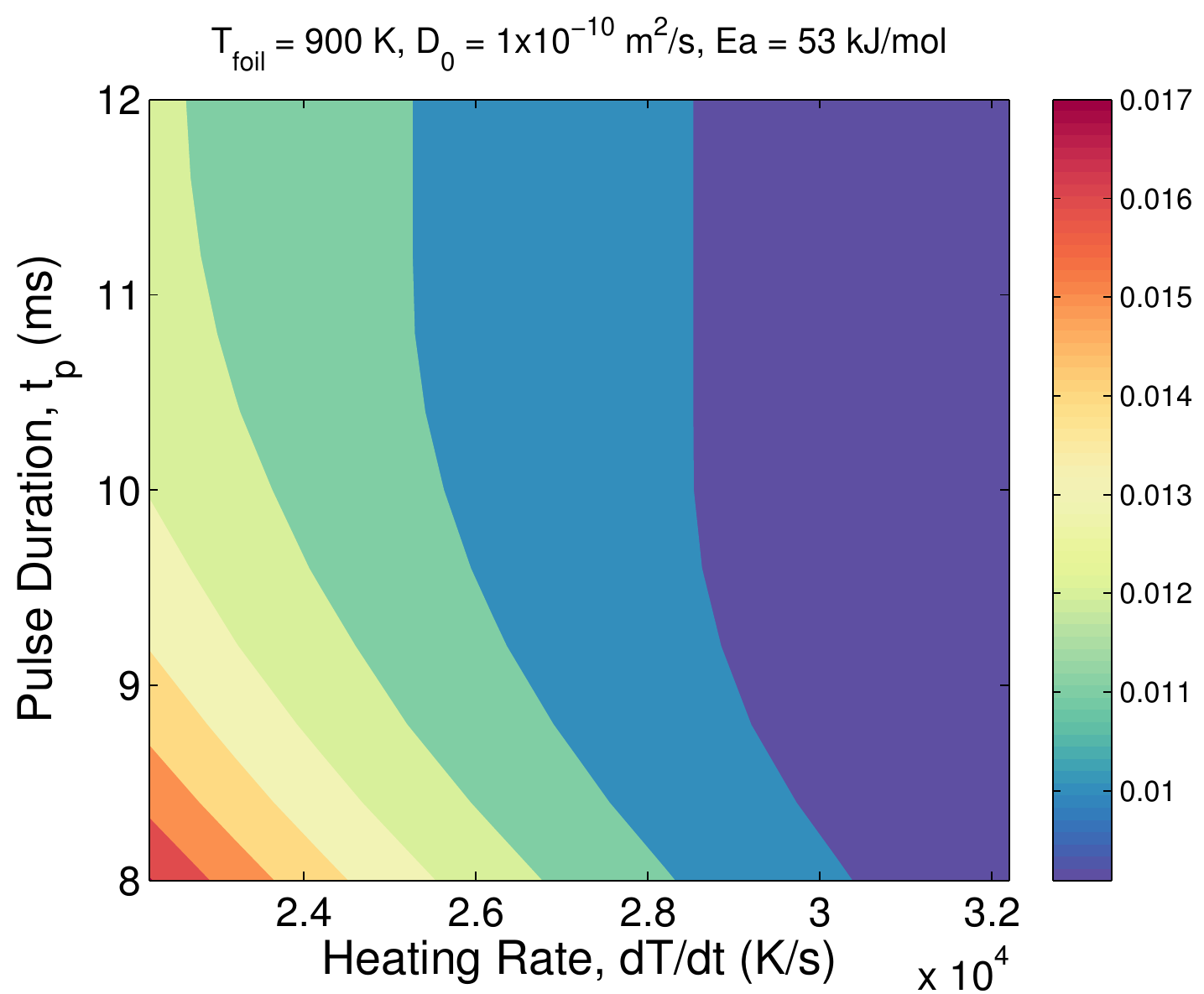}
&
\includegraphics[width=0.5\textwidth]{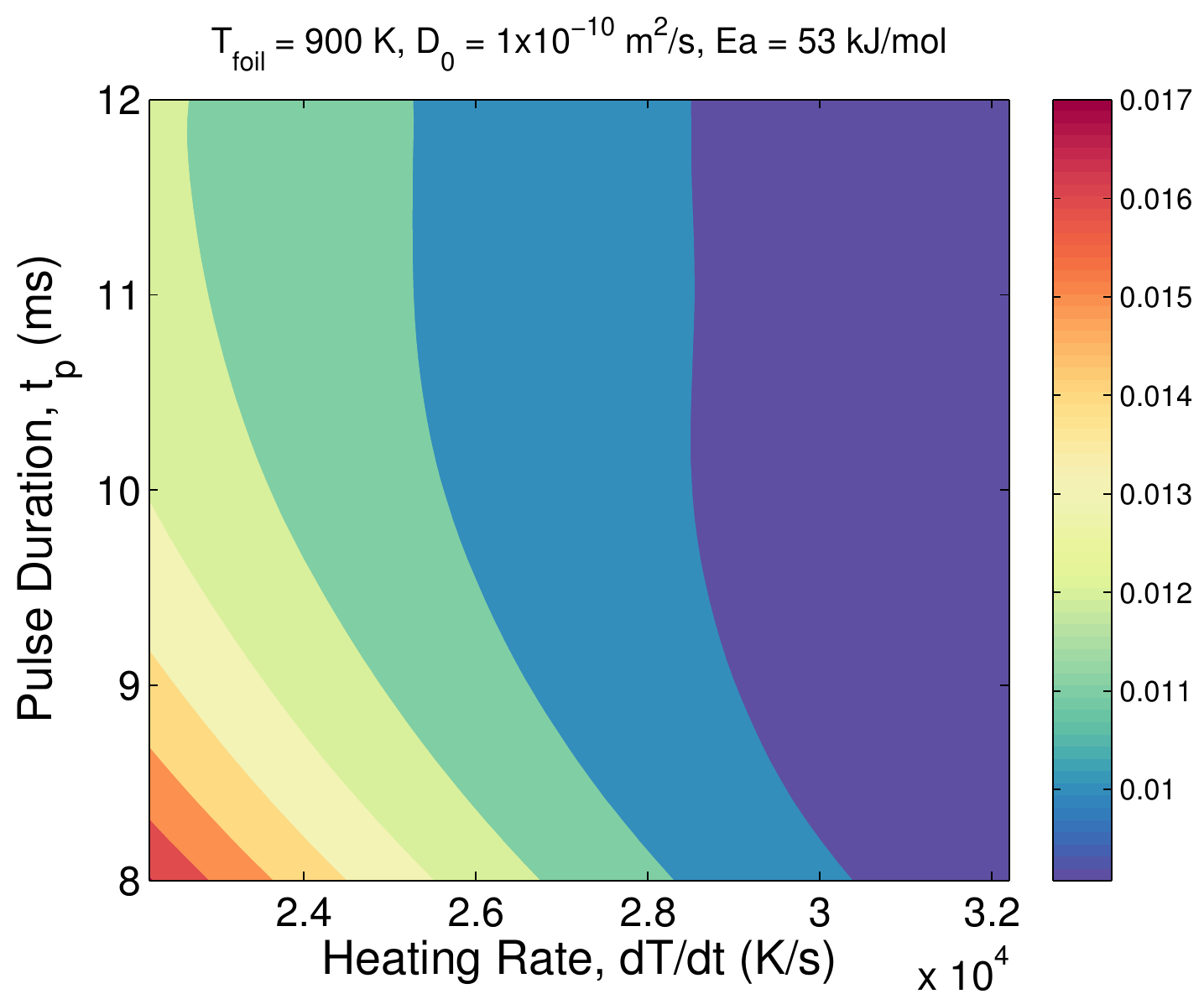}
\\ \vspace{2mm}  (a) Model & (b) PC Surrogate
\end{tabular}
\caption{Contours of reaction time as function of the physical parameters $D_{0}$ and $E_{a}$ (top) and 
of the design variables $t_{p}$ and $dT/dt$ (bottom).  The contours are generated using (a) the reaction model,
and (b) the PC surrogate.}
\label{fig:contours1}
\end{center}
\end{figure}

Figure~\ref{fig:contours1} illustrates 2D contours for the reaction time, $t_r$. 
Top row figures are generated in the $D_0$-$E_a$ plane for fixed values of the
design parameters, and bottom row figures are generated in the
$dT/dt$-$t_p$ plane using fixed values of the uncertain parameters.
The plots in the left column are generated directly from the computed
realizations, whereas contours presented on the right column are
generated using the corresponding PC representation, on a uniform Cartesian grid.
Several observations can be
made. First, a close agreement is observed between the left and the right plots,
further verifying that the PC representation is indeed suitable and
that the adapted sparse grid is sufficiently dense to capture the
response of the reaction time.  Second, $t_r$ varies more rapidly
along the $E_{a}$ axis compared to the $D_{0}$ axis.  Hence, model
predictions are expected to be more sensitive to the activation
energy.  Third, variation in $t_r$ is more pronounced as the heating rate
is varied than when the pulse duration is varied. This suggests that
the reaction time is more sensitive to $dT/dt$ than $t_p$.  Finally,
as the foil heating rate increases, the variations of the $t_r$ with
pulse duration appear to be suppressed. Similarly, as the pulse
duration increases, the variation of the reaction time with heating
rate becomes milder.  Thus, $t_r$ is expected to be more sensitive to
design conditions characterized by small heating rates and pulse
widths.

\begin{figure}[htbp]
 \begin{center}
\includegraphics[width=0.70\textwidth]{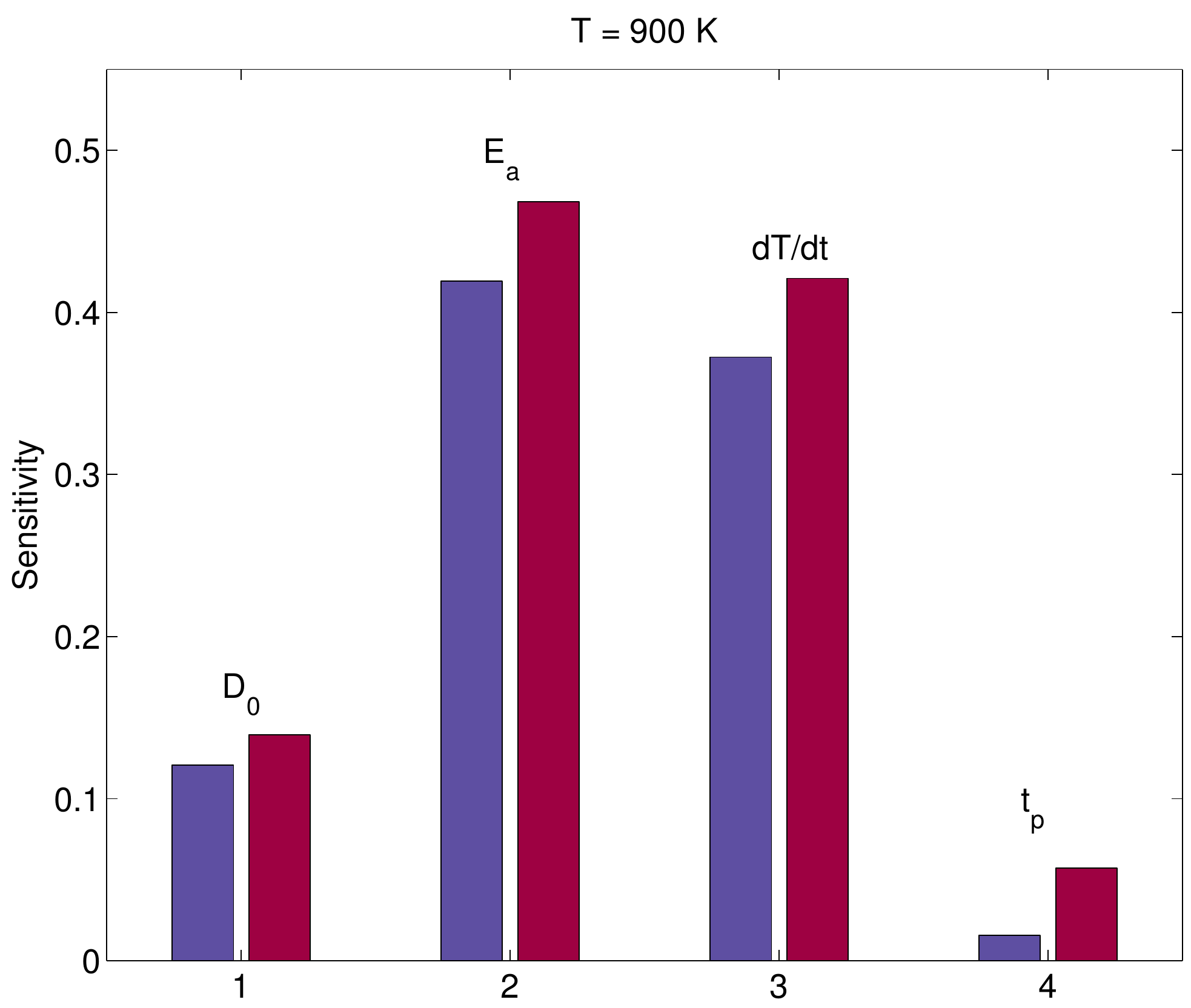}
\caption{First order (blue) and total sensitivity (red) indices for the uncertain diffusivity parameters and design variables.} 
\label{fig:sens1}
\end{center}
\end{figure}

To quantify the variability induced by the design and physical
parameters, we exploit the PC surrogates to estimate the associated
Sobol sensitivity indices.  Results are reported in
Figure~\ref{fig:sens1}, which depicts the first order and total
sensitivity indices associated with $D_{0}$, $E_{a}$,
$dT/dt$ and $t_{p}$, based on the PC expansion for the 900~K reaction
time. 

It is
observed that the model predictions are predominantly sensitive to
$E_{a}$ and $dT/dt$. This is consistent with the observations from
Figure~\ref{fig:contours1}, as well as the analysis
in~\cite{Vohra:2014b} which considered the impact of $D_0$ and $E_a$
at fixed values of $t_p$ and $dT/dt$.  In addition, we note that
except for $t_{p}$, the first and total order sensitivity indices are
close to each other indicating that the contributions due to mixed
terms (joint effects) in the PC expansion are small.  The observed
trends underscore the importance of properly selecting the design
conditions for the purpose of calibrating the model parameters, as
discussed further below.

\begin{figure}[htbp]
 \begin{center}
\includegraphics[width=0.6\textwidth]{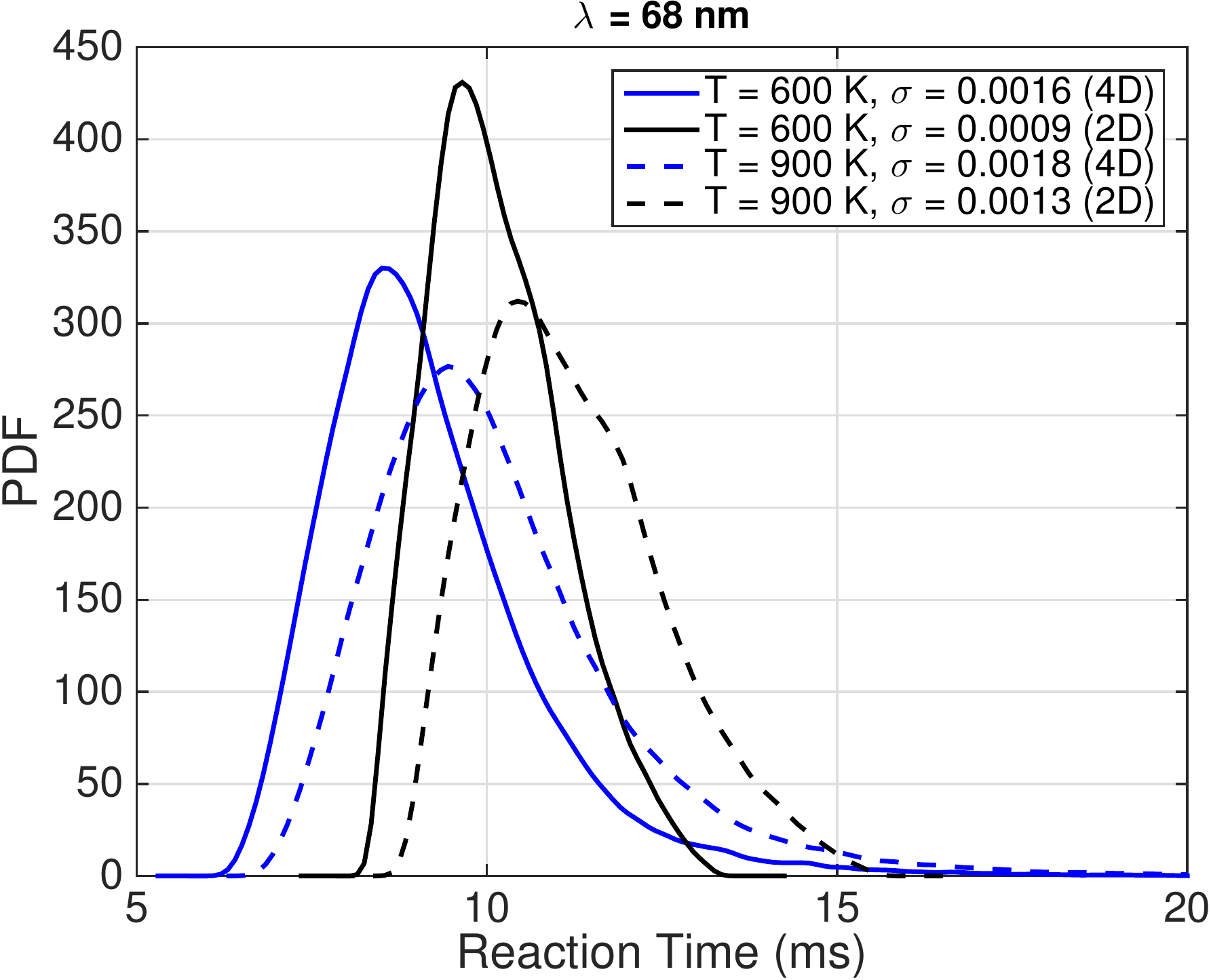}
%
\caption{Probability density functions of the reaction time, generated using
kernel density estimation.   Plotted are curves generated for a multilayer with 
$\lambda = 68$~nm using the 4D and 2D surrogates.  The 2D surrogates
are obtained from the 4D PC expansion by fixing the heating rate and 
pulse duration, respectively $t_p = 10$~ms and $dT/dt = 22200$~K/s.
For each curve, the value of the standard deviation is indicated in the label.}
\label{fig:forward1}
\end{center}
\end{figure}

Additional insight into the variability induced by the parameters is
gained by plotting PDFs of the reaction time.  The PDFs are created by
first generating a large number ($10^5$ -- $10^6$) of MC samples of
$\bm{\xi}$, evaluate the PC expansion at those points, and employ
kernel density estimation on the outputs to create smooth density
visualizations.  Figure~\ref{fig:forward1} depicts the PDFs for
reaction time due to the uncertainty jointly over parameter and design
spaces (4D case), and also for due to the uncertainty from only the
parameter space and at a fixed setting of the heating rate and pulse
duration (2D case). Plotted are curves generated based on PC representation
of the 600K and 900K reaction times.
As expected, the distributions corresponding to
the 4D case (blue curves) exhibit a wider spread (variability) as
compared to the 2D case (black curves). For both the 4D and 2D cases,
the maximum likelihood estimate (MLE) of 900K reaction time is shifted
to the right with respect to the MLE of the 600K reaction reaction
time.  Additionally, the standard deviation ($\sigma$) estimates indicate that
the coefficients of variation associated with the distributions are
large, corresponding to several fold variability in the reaction
times.

\begin{figure}[htbp]
 \begin{center}
 \begin{tabular}{cc}
 \hspace{-12mm}
\includegraphics[width=0.52\textwidth]{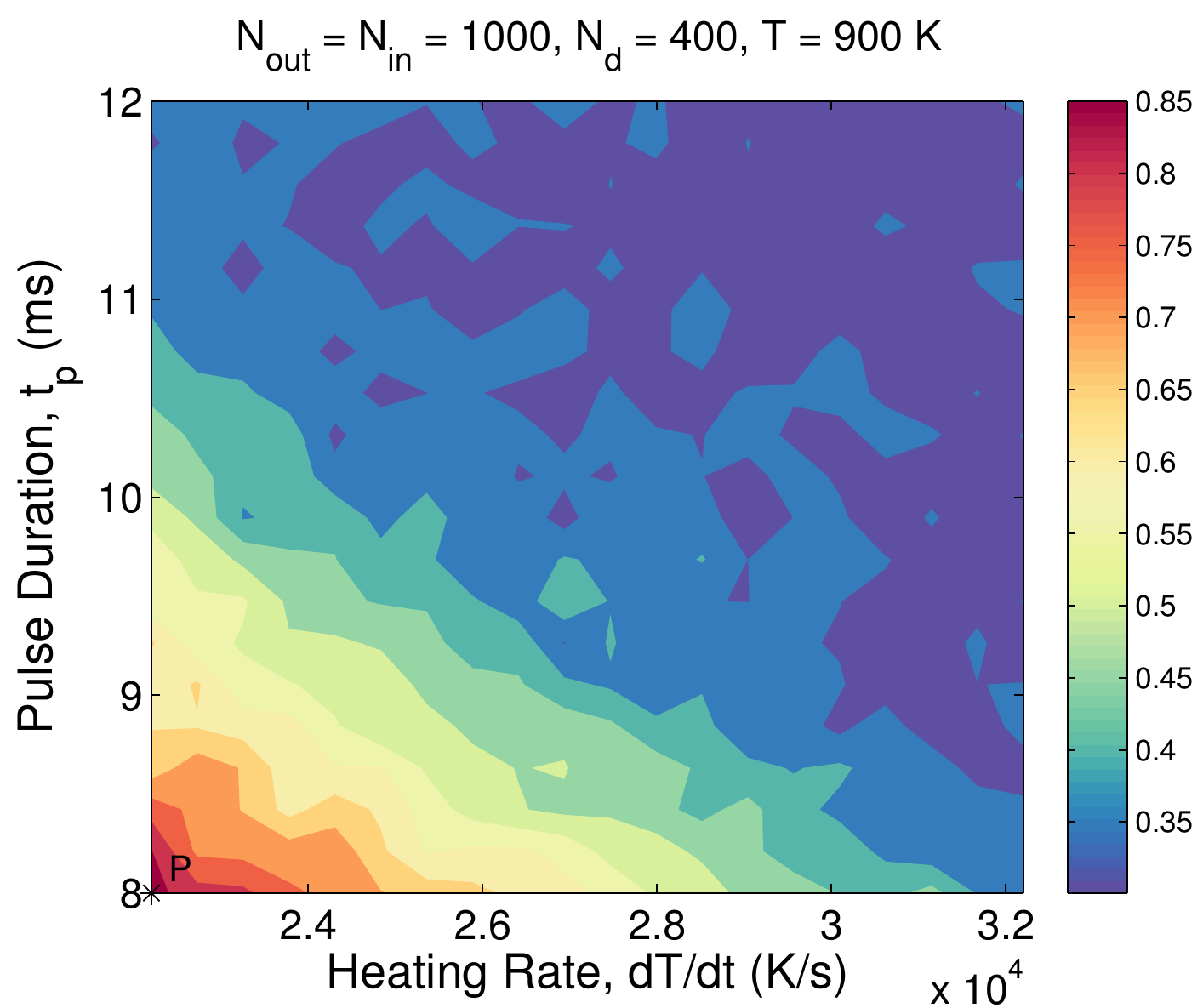}
&
\includegraphics[width=0.52\textwidth]{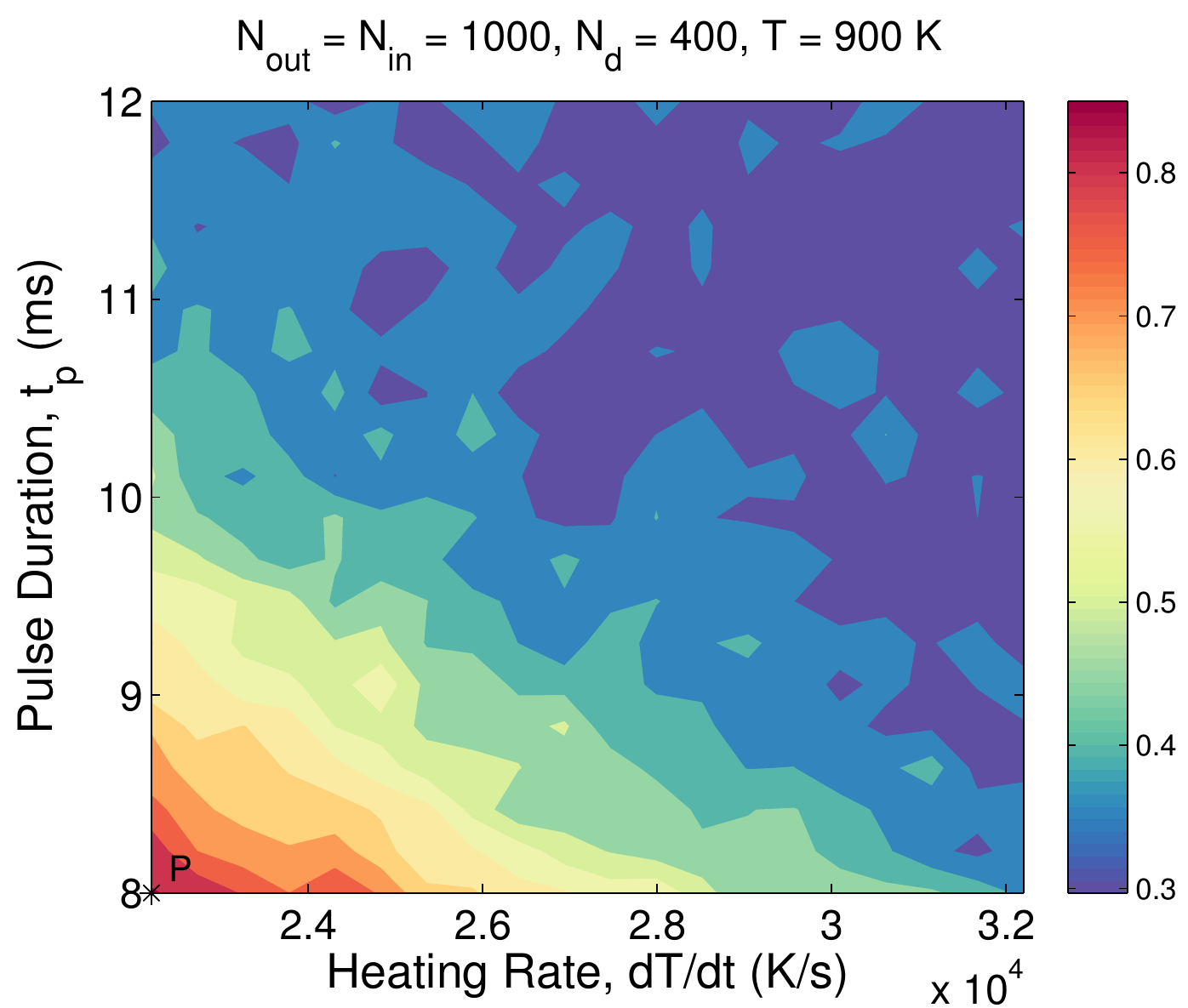}
\\ \vspace{2mm} (a) Monte Carlo Sampling & (b) Latin Hypercube Sampling
\end{tabular}
\caption{Contours of the expected utility in the considered ranges of the design variables.
Results are obtained using (a) Monte Carlo sampling and (b) Latin Hypercube sampling.  
In each case, the point \textbf{P} corresponds to the maximum utility value.}
\label{fig:utility_sam}
\end{center}
\end{figure}

We now explore the behavior of the expected utility
function~\eqref{eqn:utility}.  In this work, we perform the
optimization using a simple grid search; more sophisticated
optimization algorithms may be employed, such as those described
in~\cite{Huan:2014}, but we do not pursue them here.  The 2D design
space is discretized using a $20\times20$ uniform grid, for a total of
$N_{d}=400$ design points. At each design point, we compute a
numerical approximation of $U(\bm{d})$ via~\eqref{eqn:outer}
and~\eqref{eqn:inner}, using $N_{out} = N_{in} = 1000$ samples of the
diffusion parameters.  To generate the $\bm{y}$ samples, we first
obtain our model evaluations $\bm{M}$ by evaluating the PC surrogate
(note that the PC surrogate has replaced the expensive ODE models),
and then substitute them into~\eqref{eqn:disc}. The value of $\alpha$
was chosen to be 0.1 for the present exercise.
Figure~\ref{fig:utility_sam} illustrates contours of the expected
utility generated using surrogates constructed at a reaction
temperature of 900~K and a bilayer thickness of 72~nm.  
Results obtained using MC sampling and Latin Hypercube Sampling (LHS) agree
well with each other. The design point, \textbf{P}, that maximizes the
expected utility is located near the lower-left corner of the design
space.  In other words, experiments conducted at lower values of the
heating rate and pulse duration are anticipated to be most
informative.  The expected utility function also drops rapidly as the
pulse duration and heating rate increase, reaching a minimum near the
top-right of the design space.

\begin{figure}[htbp]
 \begin{center}
 \begin{tabular}{cc}
 \hspace{-12mm}
\includegraphics[width=0.52\textwidth]{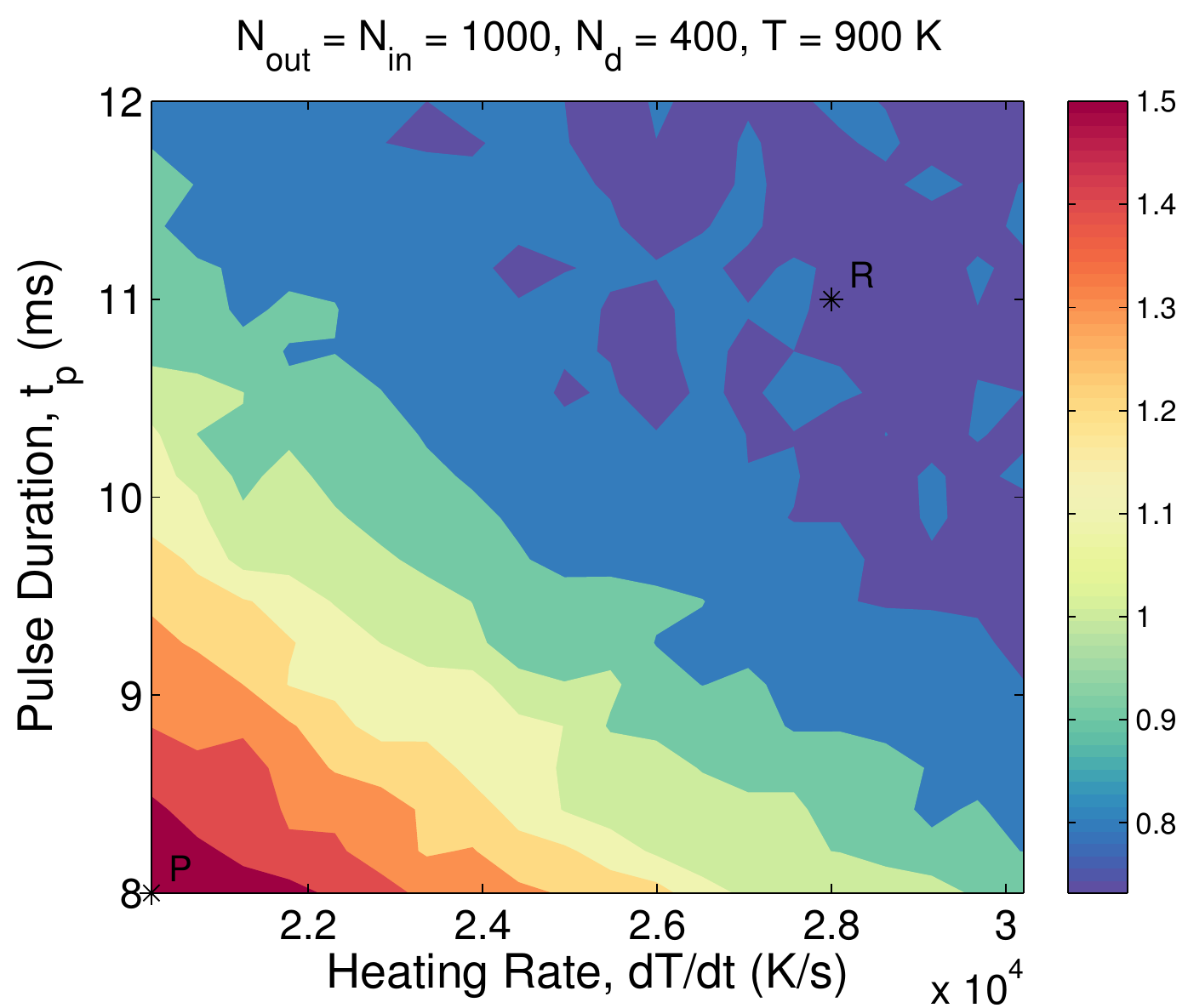}
&
\includegraphics[width=0.52\textwidth]{Ud_1000phy_400des_900K_mc_72nm.pdf}
\\ \vspace{2mm} (a) $\lambda$ = 68 nm & (b) $\lambda$ = 72 nm
\end{tabular}
\caption{Contours of the expected utility for (a) $\lambda = 68$~nm, and (b) $\lambda = 72$~nm. 
In (a), the points \textbf{P} and \textbf{R}
correspond to the maximum and minimum utility values, respectively.}
\label{fig:utility_lam}
\end{center}
\end{figure}

Figure~\ref{fig:utility_lam} contrasts utility function contours for
Zr-Al multilayers with bilayer thicknesses, $\lambda = 68$~nm and
72~nm.  Contours in both cases are qualitatively similar, and the more
informative designs reside in regions of smaller values of heating
rate and pulse duration.  However, quantitative differences can be
observed.  These are indeed expected, as the reaction times are
expected to depend significantly on the bilayer
thickness~\cite{Fritz:2013,Fritz:2011,Morris:2010,Spey:2006}.
Highlighted in Figure~\ref{fig:utility_lam}(a) are two extreme
scenarios corresponding to the maximum and minimum utility
estimates. Design conditions at point \textbf{P} are regarded as
optimal, whereas conditions at point \textbf{R} are expected to be
least informative.  The ratio of the expected utility estimated for
the design \textbf{P} to that of design \textbf{R} is approximately
3.08 for the $\lambda$ = 72 nm case, and approximately 2 for the
$\lambda$ = 68 nm case. Hence, the information gain from the posterior
to the prior is expected to be much higher when data from optimally
designed experiments is used for parameter inference.

\begin{figure}[htbp]
 \begin{center}
 \begin{tabular}{cc}
 \hspace{-12mm}
\includegraphics[width=0.52\textwidth]{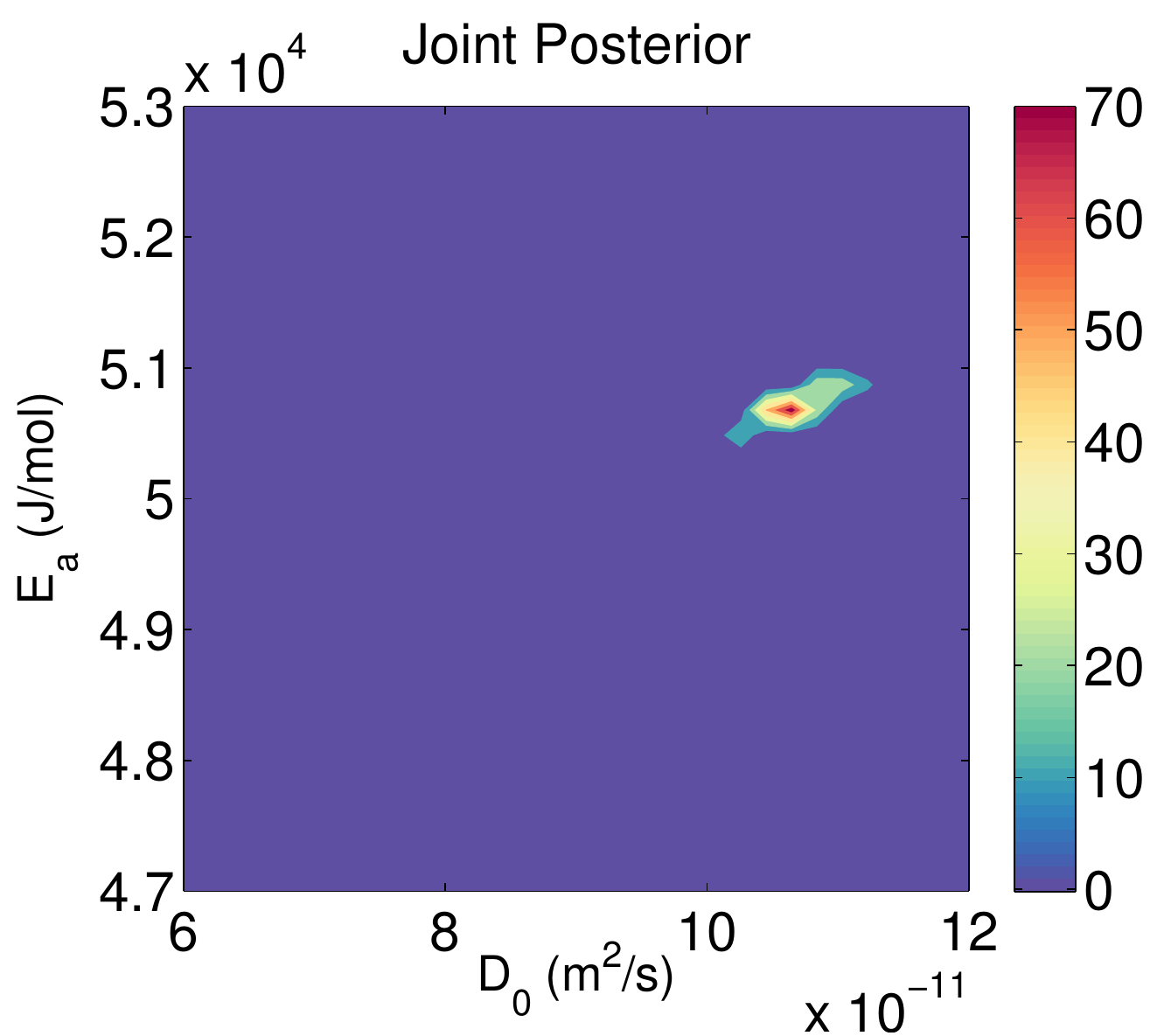}
&
\includegraphics[width=0.52\textwidth]{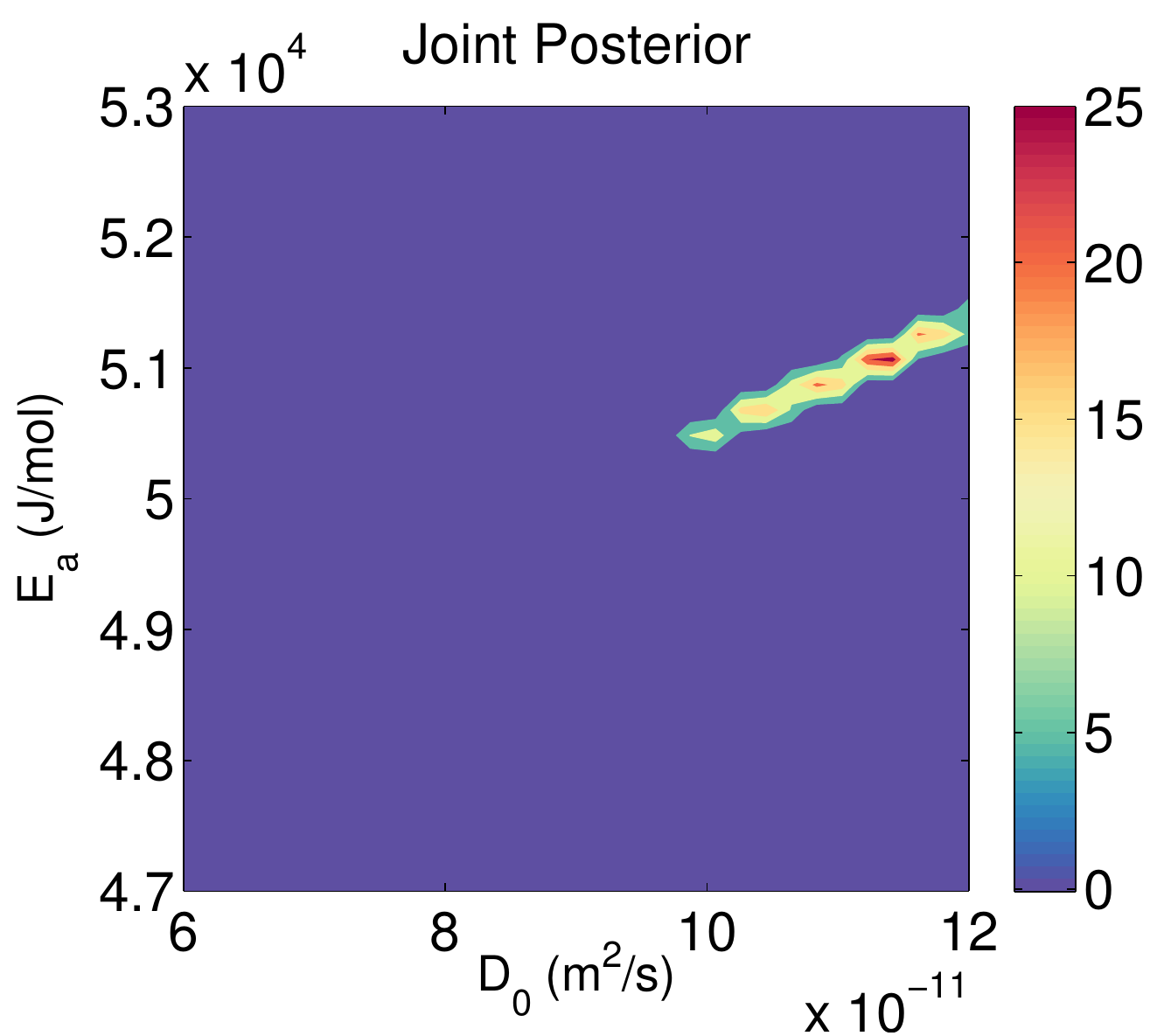}
\\ \vspace{2mm} (a) \textbf{P} & (b) \textbf{R}
\end{tabular}
\caption{Joint posterior distribution of $D_{0}$ and $E_{a}$.
Contours are generated for design conditions corresponding to 
(a) point \textbf{P} and (b) point \textbf{R} in Fig.~\ref{fig:utility_lam}(a).}
\label{fig:post_PR}
\end{center}
\end{figure}

To validate the predictions based on the utility function, a Bayesian
inference exercise is conducted based on synthetic data, generated by
taking $D_{0} = 1 \times 10^{-10}$~m$^2$/s and $E_a = 51$~kJ/mol as
the \textit{``truth''} parameter values. 
To generate the synthetic observations, we first evaluate
the ODE model for the reaction time, $t_{r}$ at foil temperatures
ranging from 700~K to 900~K, at increments of 5~K.  The resulting
reaction time estimates are then substituted into~\eqref{eqn:disc},
and corrupted by a Gaussian noise ($\alpha$ = 0.01) to generate the
set of experimental observations, $\bm y$.  Joint posterior
distributions of the diffusion parameters based on the synthetic
observations are illustrated in Figure~\ref{fig:post_PR}, for the two
designs labeled in Figure~\ref{fig:utility_lam}(a). 
The inference
results are consistent with the expected utility: the results in
Figure~\ref{fig:post_PR} indicate a more concentrated (sharper)
posterior distribution for design \textbf{P} than for \textbf{R}.
However, in both cases, a small bias from the assumed truth is
observed.  This is expected since observations are noisy, and also
because surrogates are used in lieu of full model computations in both
the experimental design and inference phases.

\begin{figure}[htbp]
 \begin{center}
 \begin{tabular}{cc}
 \hspace{-12mm}
\includegraphics[width=0.52\textwidth]{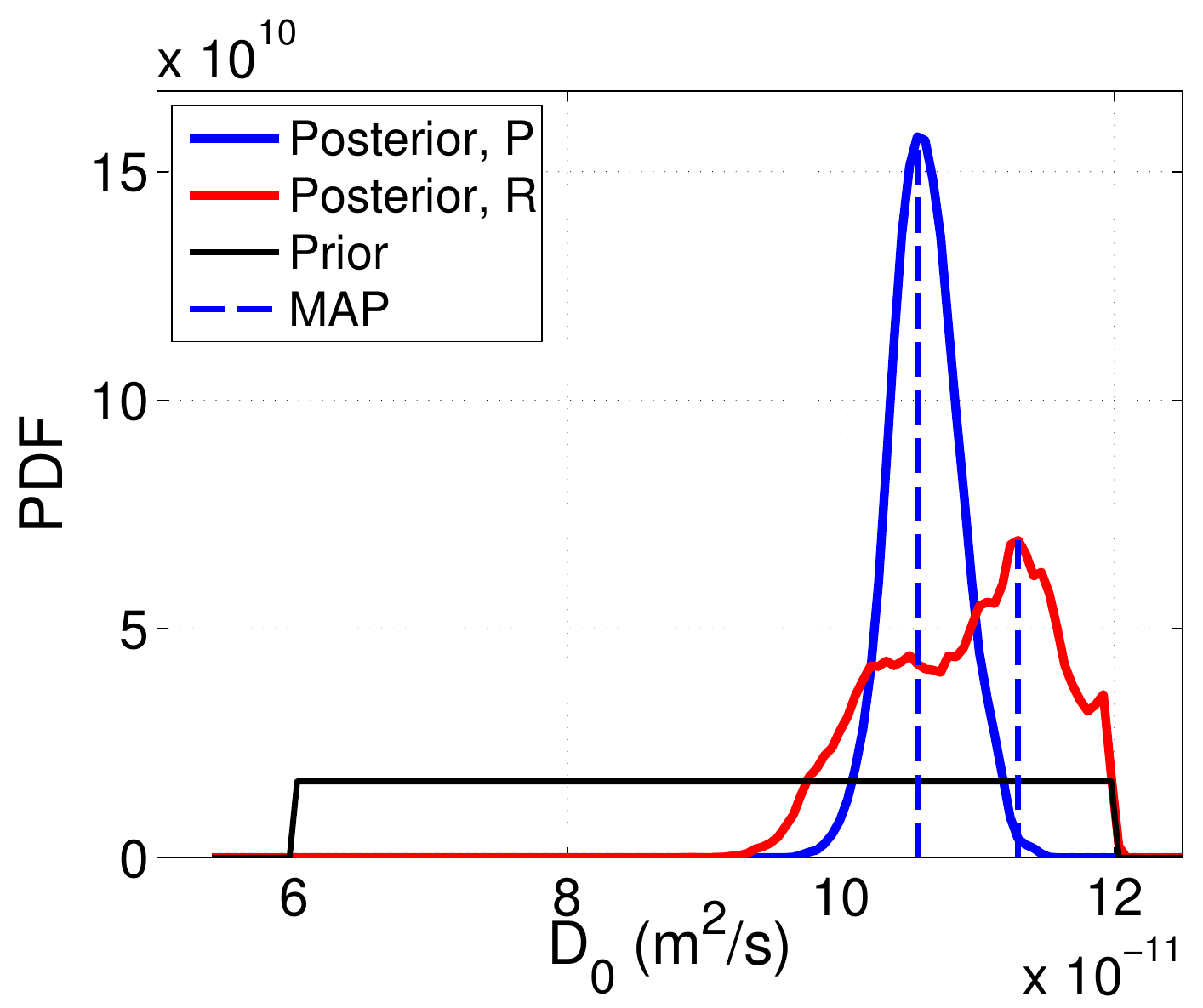}
&
\includegraphics[width=0.52\textwidth]{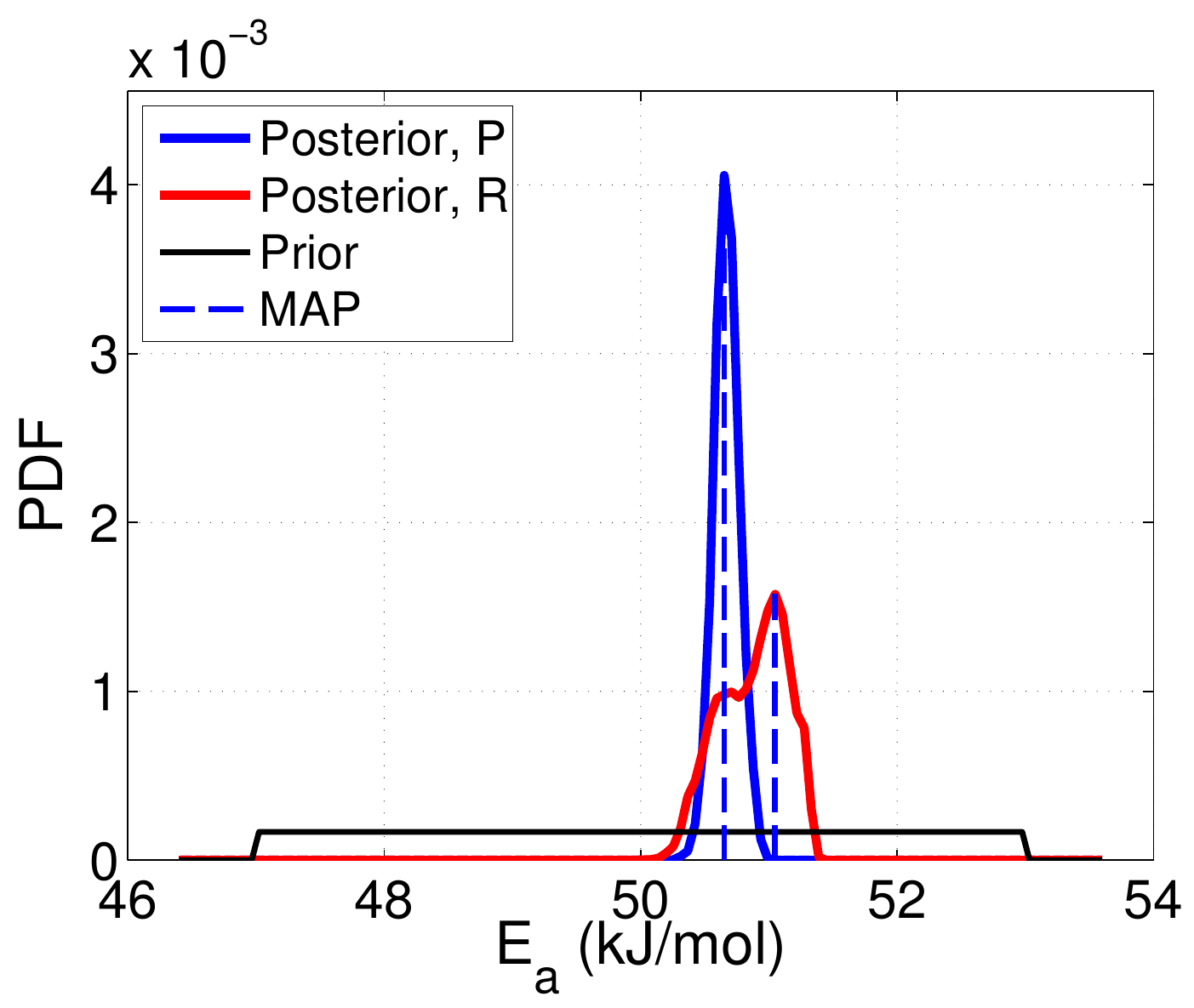}
\\ \vspace{2mm} (a)  & (b) 
\end{tabular}
\caption{Marginal distribution of the posterior of (a) $D_{0}$ and (b) $E_{a}$.  Curves are generated for design conditions corresponding
to points \textbf{P} and \textbf{R} in Fig.~\ref{fig:utility_lam}(a).  The uniform priors and the MAP estimates are also plotted.}
\label{fig:marg_PR}
\end{center}
\end{figure}

To further investigate the inference results, we compare in
Figure~\ref{fig:marg_PR}, the posterior marginals for $D_{0}$ and
$E_{a}$ corresponding to the design conditions \textbf{P} and
\textbf{R}.  We observe that posterior marginals corresponding to
design conditions \textbf{P} are much more concentrated than those
corresponding to point \textbf{R}, again indicating that the former
are indeed more informative.  In addition, the maximum \textit{a
  posteriori} (MAP) estimates in the case of the optimal design
conditions are in better agreement with the assumed truth
parameters.

\subsection{Self-Propagating Reactions}
\label{ssec:prop}

For self-propagating front experiments, we again seek design
conditions that maximize the expected information gain on $D_0$ and
$E_0$. The design variables are now the initial foil temperature,
$T_0$, and premix width, $w_0$. The parameters are assumed to have
uniform priors in the following ranges: $D_{0} \in [1.5\times 10^{-9},
  4.5\times 10^{-9}]$~m$^{2}$/s and $E_{a} \in [48, 62]$~kJ/mol. The
design variables are assigned uniform measures: $T_{0} \in [300,
  500]$~K and $w_{0} \in [0.5, 1.5]$~nm. Our observable/QoI is the
front velocity, $v$.  Table~\ref{t:all_variables} summarizes all the
variables.

In this section, we apply a simplified
approach to the design of self-propagating front experiments.  The
application of a simplified methodology is motivated by the fact that
the front velocity is the main observable, and so a single experiment
(performed using a nanolaminate foil of a specified bilayer thickness)
 yields a single data point only.  One is thus faced with
the challenge
of dealing with a limited observation set, possibly involving noisy
data.  Several approaches have been applied to address this challenge,
including performing repeated measurements at the same conditions,
namely to quantify the impact of measurement uncertainty.  Another
avenue is based on the variation of foil properties.  The most common
approach falling within this framework concerns conducting experiments
with different bilayer thicknesses.  Whereas this approach has proven
to be effective, it generally requires different deposition runs to
synthesize samples with the desired properties, and to characterize
the samples through imaging and calorimetry.  These synthesis and
analysis steps can be quite costly and time consuming, which leads to
inherent constraints on the experimental sample size.  An alternative
approach is to conduct experiments with the same bilayer thickness but
different premixed widths, which can be altered post fabrication
through low-temperature annealing.

In light of the considerations above, we adopt a simplified approach
that enables us to consider two different scenarios, and to draw
inferences in an efficient manner.  For each of these settings, we
rely on a Bayesian inference methodology based on synthetic, noisy
velocity data, and analyze the resulting information gain.  In the
first, we assume a design strategy in which experiments are conducted
for multilayers with bilayer thicknesses falling at uniform intervals
in a broad range.  In the second scenario, a design methodology is
assumed in which repeated measurements are conducted for samples
having disparate values of the bilayer thickness.  In addition, we
assess the impact of sample size of
the available measurements, and of measurement uncertainty on
posterior marginals of the diffusion parameters. We then contrast
posterior marginals pertaining to the two different strategies. As in
the analysis above, we rely on PC surrogates of the observable in
order to enable efficient sampling of posterior distributions. 

We then construct PC surrogates for the self-propagating front
velocity according to: \be v(\lambda) = \sum_{\bm{k}=0}^{P}
c_{\bm{k}}(\lambda)\Psi_{\bm k}(\bm{\xi}) \ee where $\bm{\xi}$ is
four-dimensional corresponding to the stochastic degrees of freedom in
$D_0$, $E_a$, $T_0$, $w_0$. The coefficients
$c_k$ are computed using the aPSP algorithm, requiring 36 iterations
and a total of 2225 independent model simulations.
The four-dimensional
polynomial basis of the resulting surrogate in this case is observed to be of the
order: 2, 2, 6 and 2 along $\xi_1$, $\xi_2$, $\xi_3$ and $\xi_4$ respectively.
Relative error is
computed for surrogates generated for different bilayer thicknesses
falling in the range 40~nm~--~90~nm, and the peak value of the relative
error is found to be less than 0.45\%. The aPSP construction thus provides
excellent representations of the reaction velocity behavior in all
cases considered in the present setup.

\begin{figure}[htbp]
 \begin{center}
 \begin{tabular}{cc}
\includegraphics[width=0.5\textwidth]{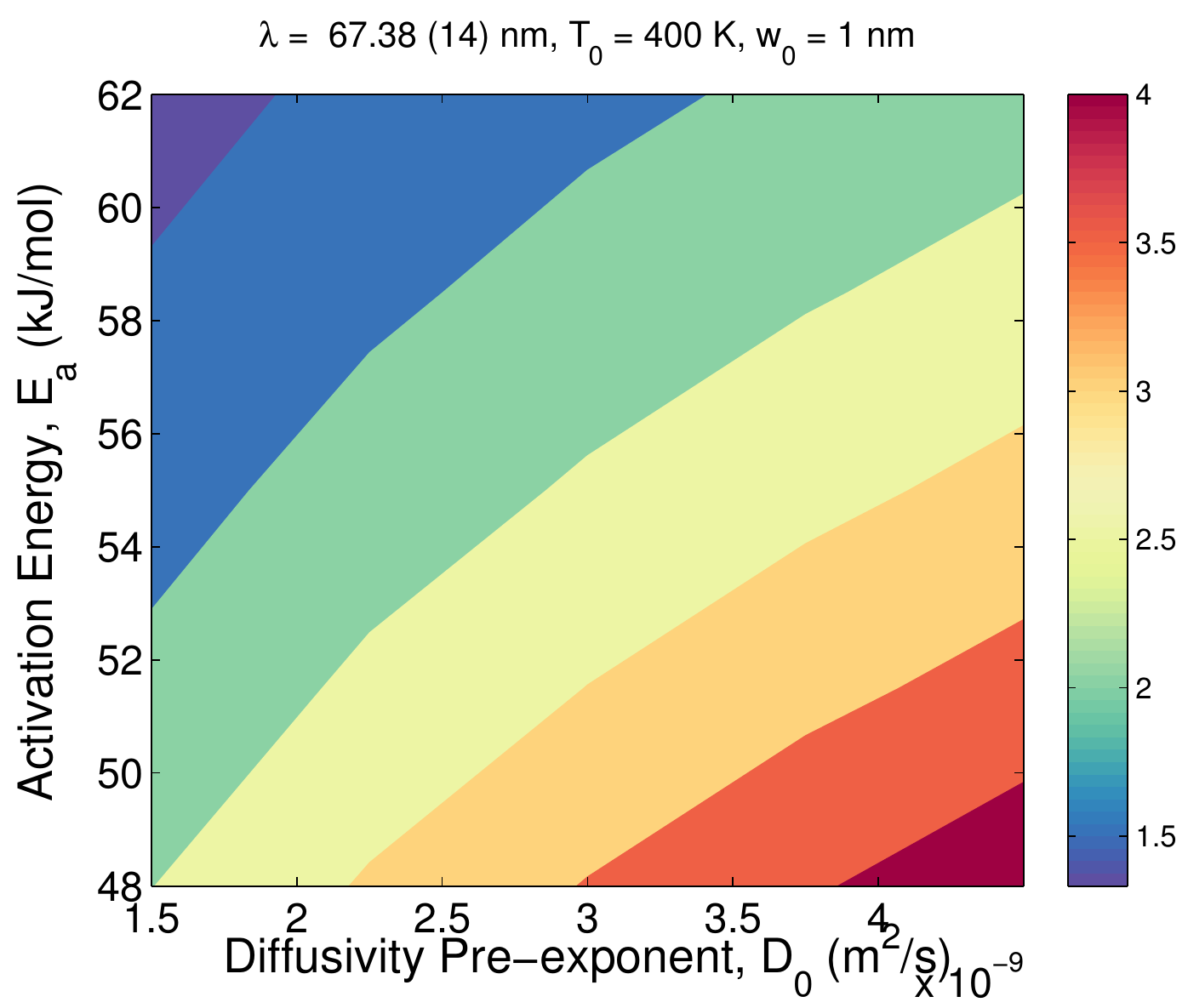}
&
\includegraphics[width=0.5\textwidth]{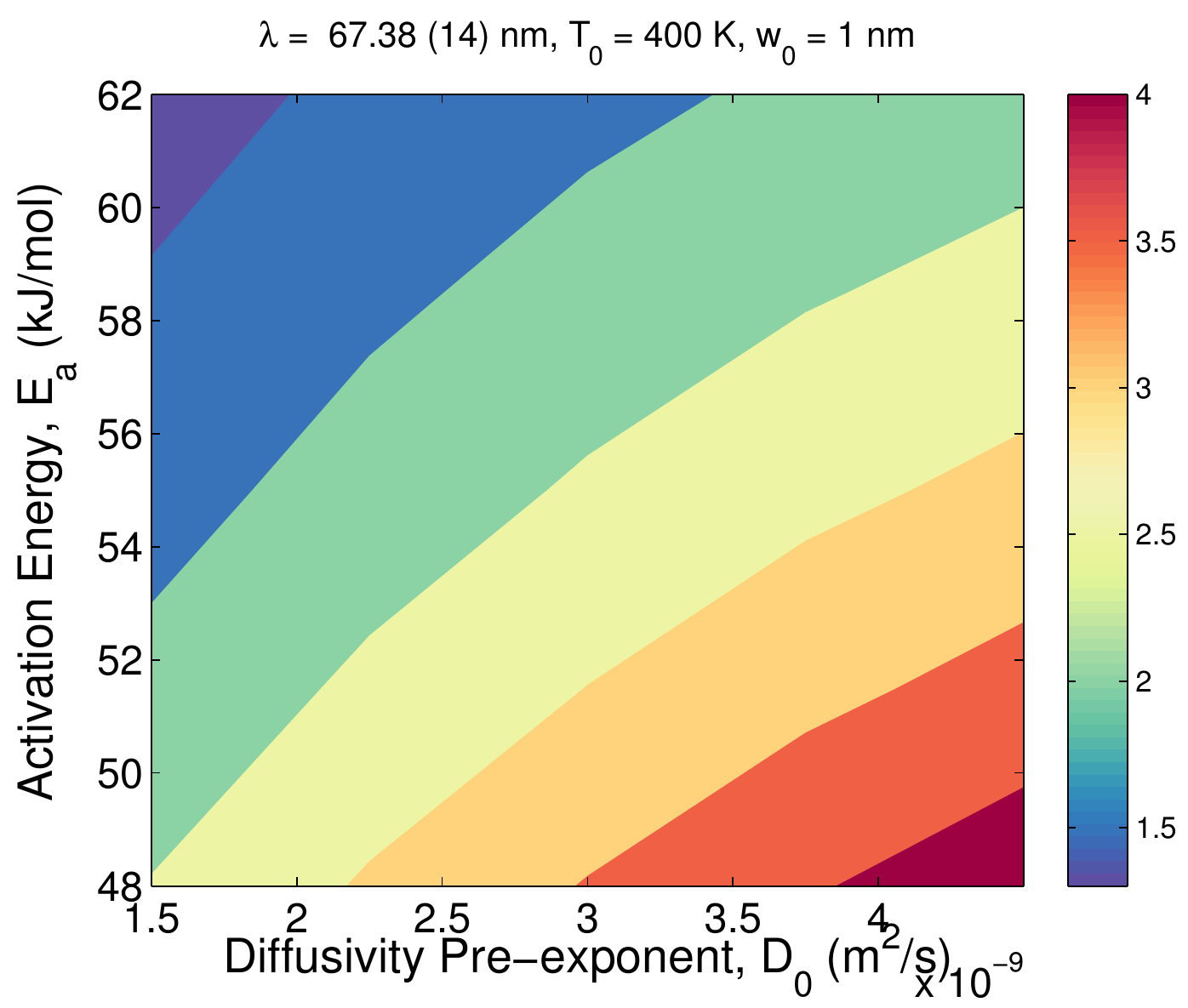}
\\
\includegraphics[width=0.5\textwidth]{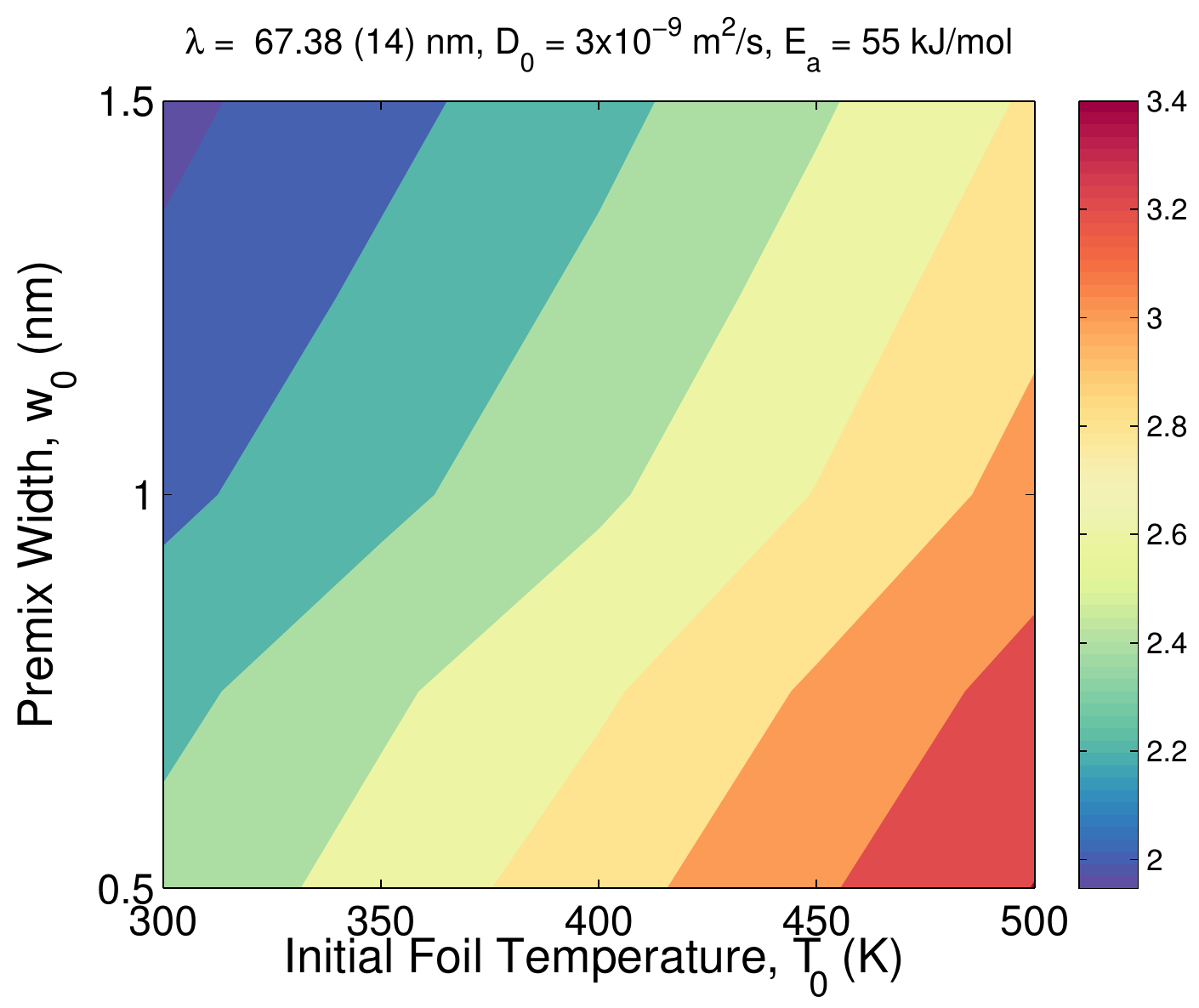}
&
\includegraphics[width=0.5\textwidth]{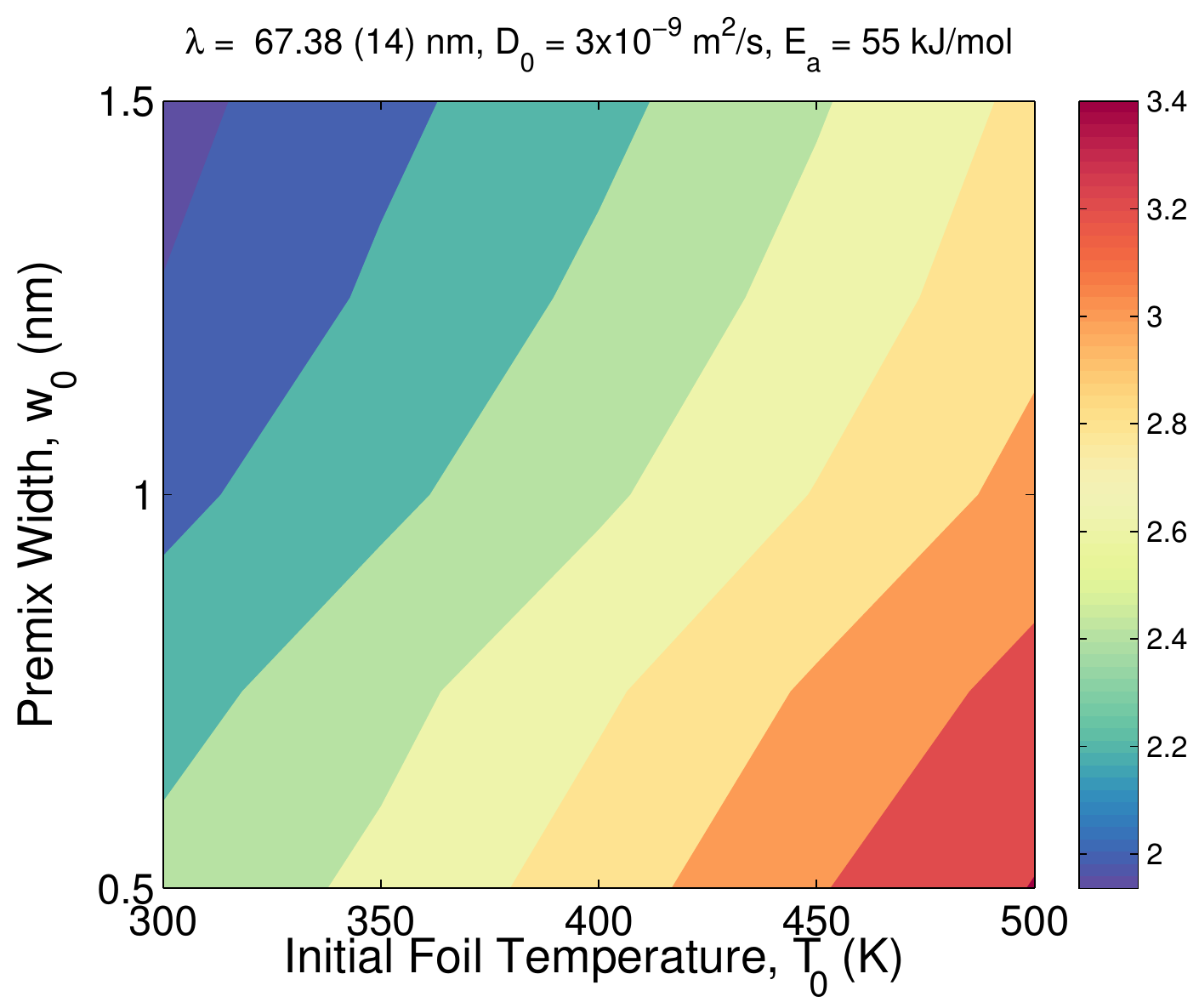}
\\ \vspace{2mm}  (a) Model & (b) PC Surrogate
\end{tabular}
\caption{Contours of reaction velocity as function of the physical parameters, $D_{0}$ and $E_{a}$ (top) and 
of the design variables, $t_{p}$ and $dT/dt$ (bottom).  The contours are generated using (a) the reaction model,
and (b) the PC surrogate.}
\label{fig:contours2}
\end{center}
\end{figure}

Figure~\ref{fig:contours2} illustrates reaction velocity contours for
a Zr-Al multilayer with bilayer thickness, $\lambda=67.38$~nm.  Plotted
are 2D contours for the considered range of diffusion parameters using
fixed values of the design parameters, and for the considered range of
design conditions with fixed values of the diffusion parameters.  The
contours are generated directly from the realizations, and by sampling
the PC representation of the front velocity on uniform grids.
Furthermore, Figure~\ref{fig:contours2} indicates that, as expected,
the reaction velocity increases as the diffusivity pre-exponent
increases, and decreases as the activation energy increases.  In
addition, the reaction velocity increases with $T_{0}$, and decreases
with $w_{0}$.  The results also show that contours generated directly
from the simulation data are in close agreement with those obtained by
sampling the PC representation.  This is consistent with the PC
relative errors.

\clearpage

\begin{figure}[htbp]
 \begin{center}
\includegraphics[width=0.7\textwidth]{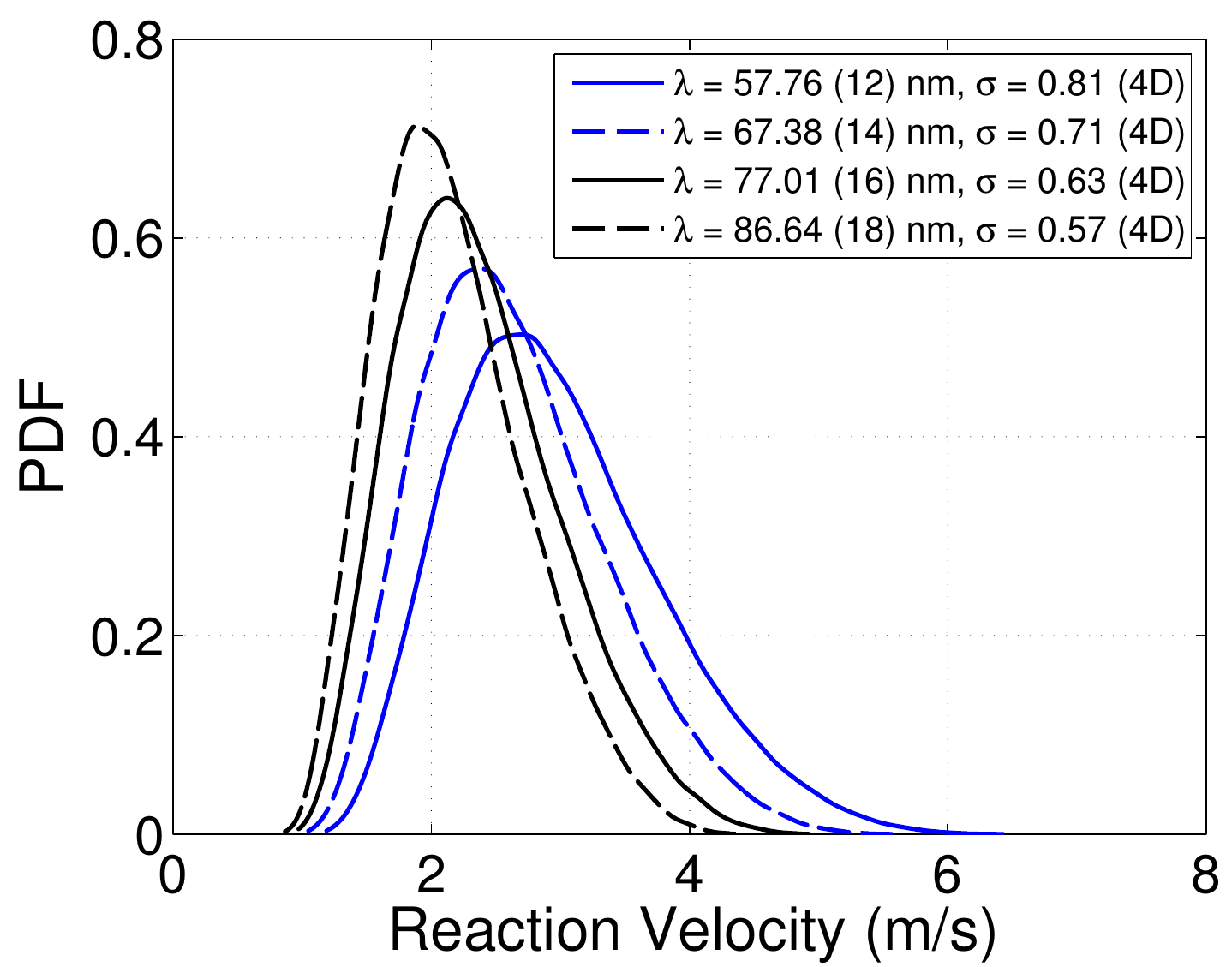}
\caption{
PDFs of reaction velocity for $\lambda = 57.76$~nm, 67.38~nm, 77.01~nm and 86.64~nm.
Reported in brackets is the corresponding thickness of the aluminum layer in each case.}
\label{fig:forward}
\end{center}
\end{figure}

Plotted in Figure~\ref{fig:forward} are PDFs of reaction velocity as a
result of uncertainty in both the parameters and design variables, and
generated using PC surrogates constructed for four different values of
$\lambda$. The predictions reveal that the peak of the
distribution shifts towards lower velocities as $\lambda$ increases.
This is consistent with previous experimental, analytical and
computational
results~\cite{Knepper:2009,Mann:1997,Besnoin:2002,Vohra:2014b}, which
indicate that the front velocity decreases with bilayer thickness, as
long as the latter is sufficiently larger than the thickness of the
premixed layer.  
The spread in the distributions, as quantified by the
standard deviation, $\sigma$, is also observed to decrease with $\lambda$.
However, for the range of parameters and bilayers considered, the
coefficient of variation, defined as the ratio of the standard
deviation to the mean, is observed to be tightly bounded in the range
$[0.2635, 0.2714]$. Thus, the variability of the front velocity over
the considered parameters domains has similar impact for the entire
range of bilayers considered.

\begin{figure}[htbp]
 \begin{center}
 \begin{tabular}{cc}
 \hspace{-12mm}
\includegraphics[width=0.52\textwidth]{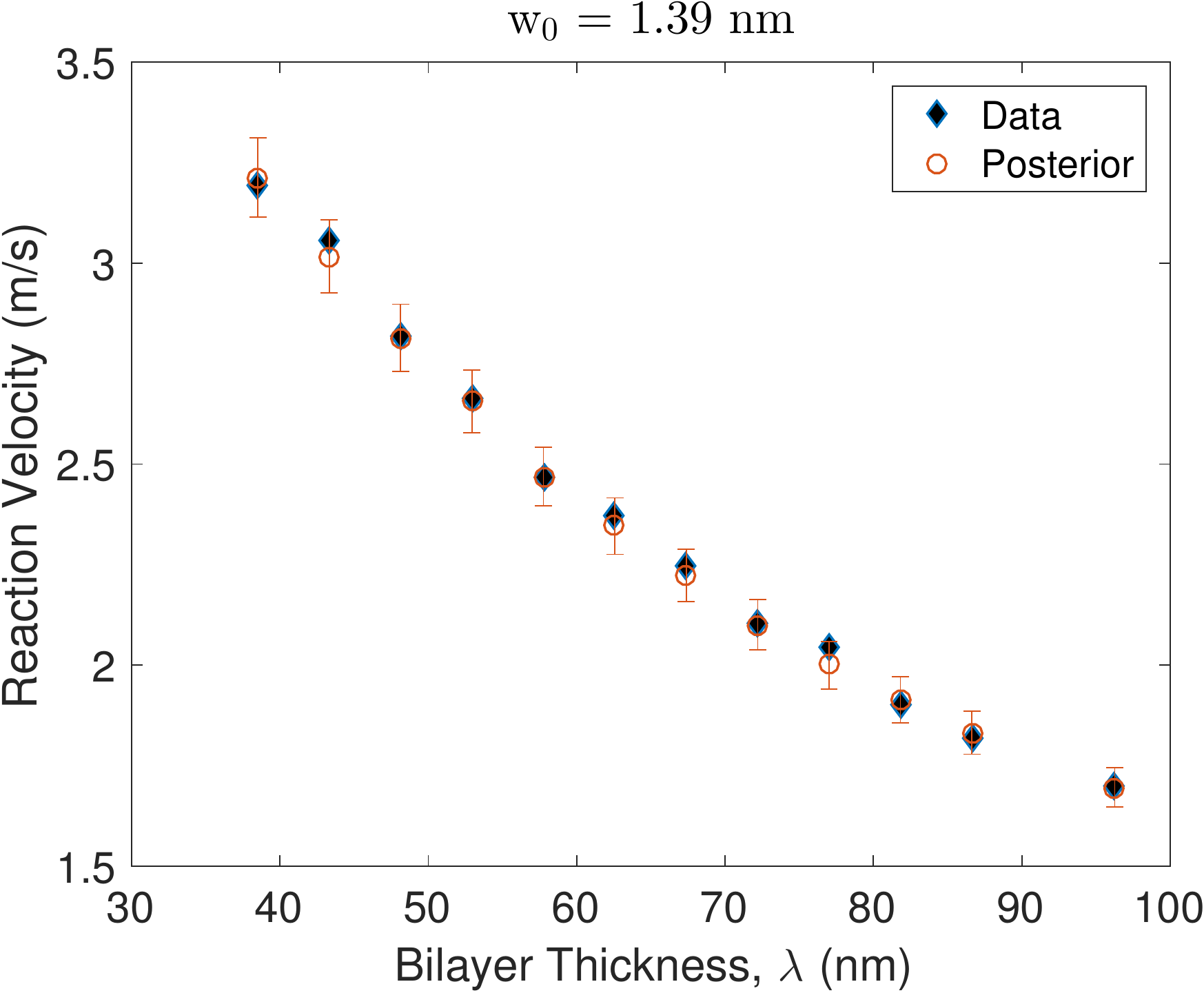}
&
\includegraphics[width=0.52\textwidth]{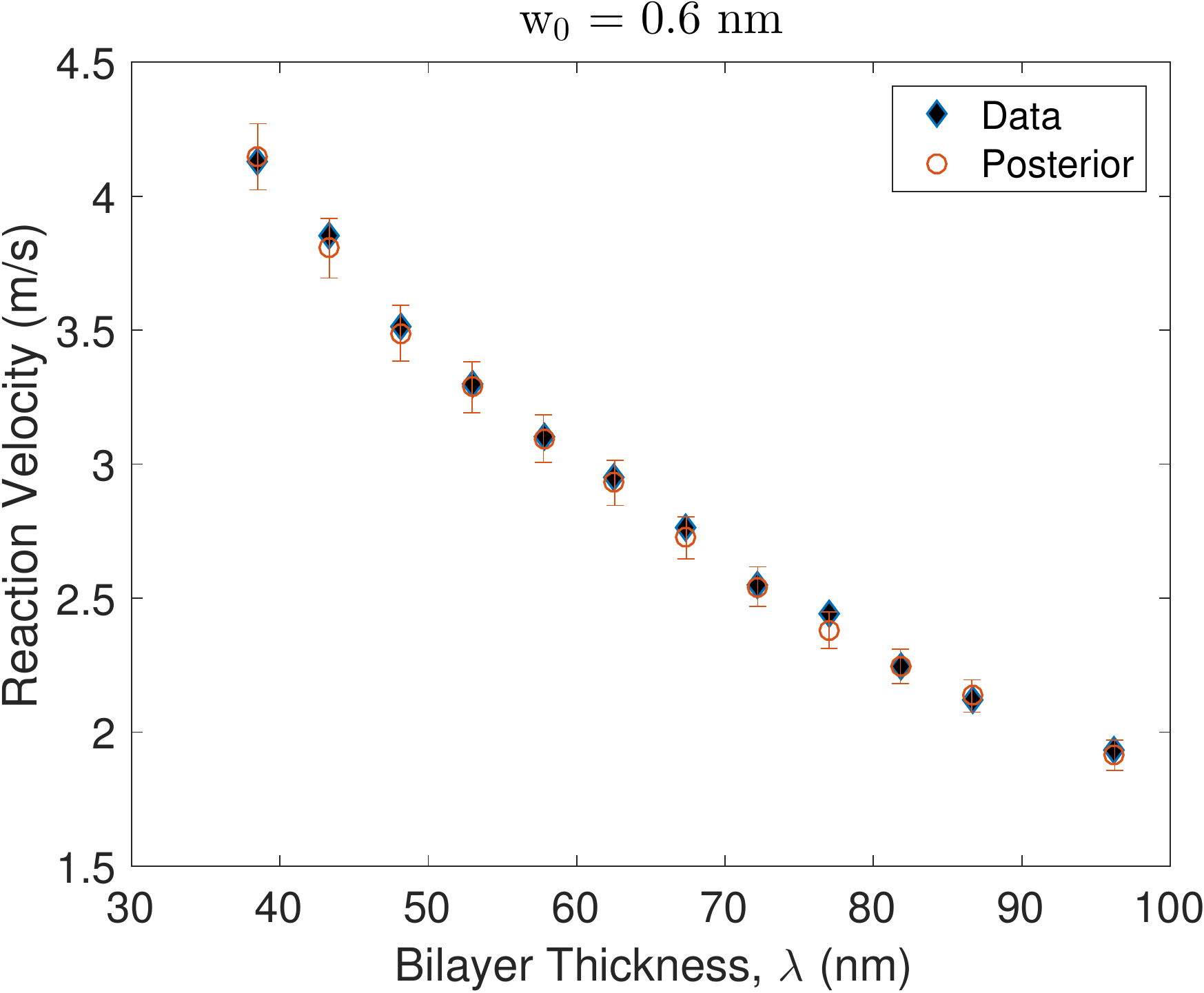}
\\ \vspace{2mm} (a)  & (b) 
\end{tabular}
\caption{Reaction velocity versus bilayer thickness for (a) $w_{0} = 1.39$~nm, and (b) $w=0.6$~nm. The synthetic
data is depicted using symbols.  The vertical bars correspond to $\pm 1$ standard deviation of the posterior predictive.}
\label{fig:ppc}
\end{center}
\end{figure}

To validate our approach involving parameter inference in a Bayesian
setting, we perform a posterior predictive check using two different
values of the premix width, $w_0= 1.39$~nm, and $w_0 = 0.6$~nm. For
each bilayer thickness considered, MCMC samples are used to determine
the mean value and standard deviation of the reaction velocity.  The
posterior predictive results are contrasted with the original data
used in the inference, as illustrated in Figure~\ref{fig:ppc}. 
 It is seen that in all cases the posterior predictive captures the data
within reasonable bounds.  This provides confidence in the inferred
diffusion parameters and corresponding velocity estimates.

\begin{figure}[htbp]
 \begin{center}
\includegraphics[width=0.6\textwidth]{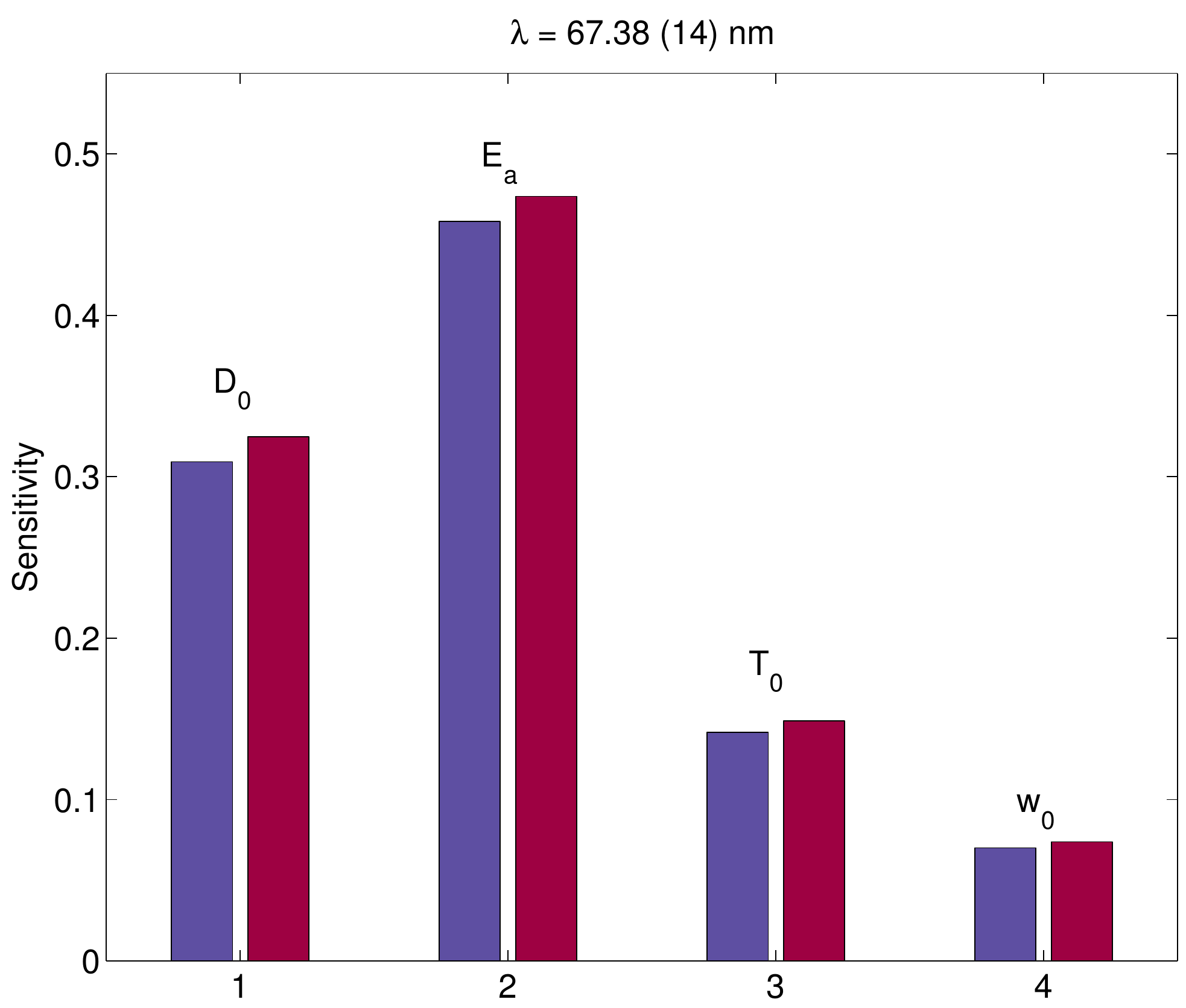}
\caption{First order (blue) and total sensitivity (red) indices for the uncertain diffusivity parameters and design variables.} 
\label{fig:sens2}
\end{center}
\end{figure}

The first and total order Sobol sensitivity indices for the
four uncertain parameters are shown using bar graphs in
Figure~\ref{fig:sens2}. 
It is seen that the model predictions for reaction velocity are most
sensitive to $E_{a}$, followed by $D_{0}$, $T_{0}$ and $w_{0}$.
Specifically, the sensitivity indices of $E_a$ and $D_0$ are large,
and so the impact of these parameters is clearly dominant.  The
sensitivity indices of $T_0$ and $w_0$ are relatively smaller, but
their impact is still significant.  
The results in
Figure~\ref{fig:sens2} also show that first-order and total
sensitivity indices of all four parameters are close to each other,
which indicates a small contribution from mixed terms to the variance
of the front velocity.  
Thus, in the parameter range considered, the
contributions of individual parameters are in large part additive.
This is not surprising because variations of the premixed width and
initial temperature are naturally restricted to relatively small
ranges.

\begin{figure}[htbp]
 \begin{center}
\includegraphics[width=0.6\textwidth]{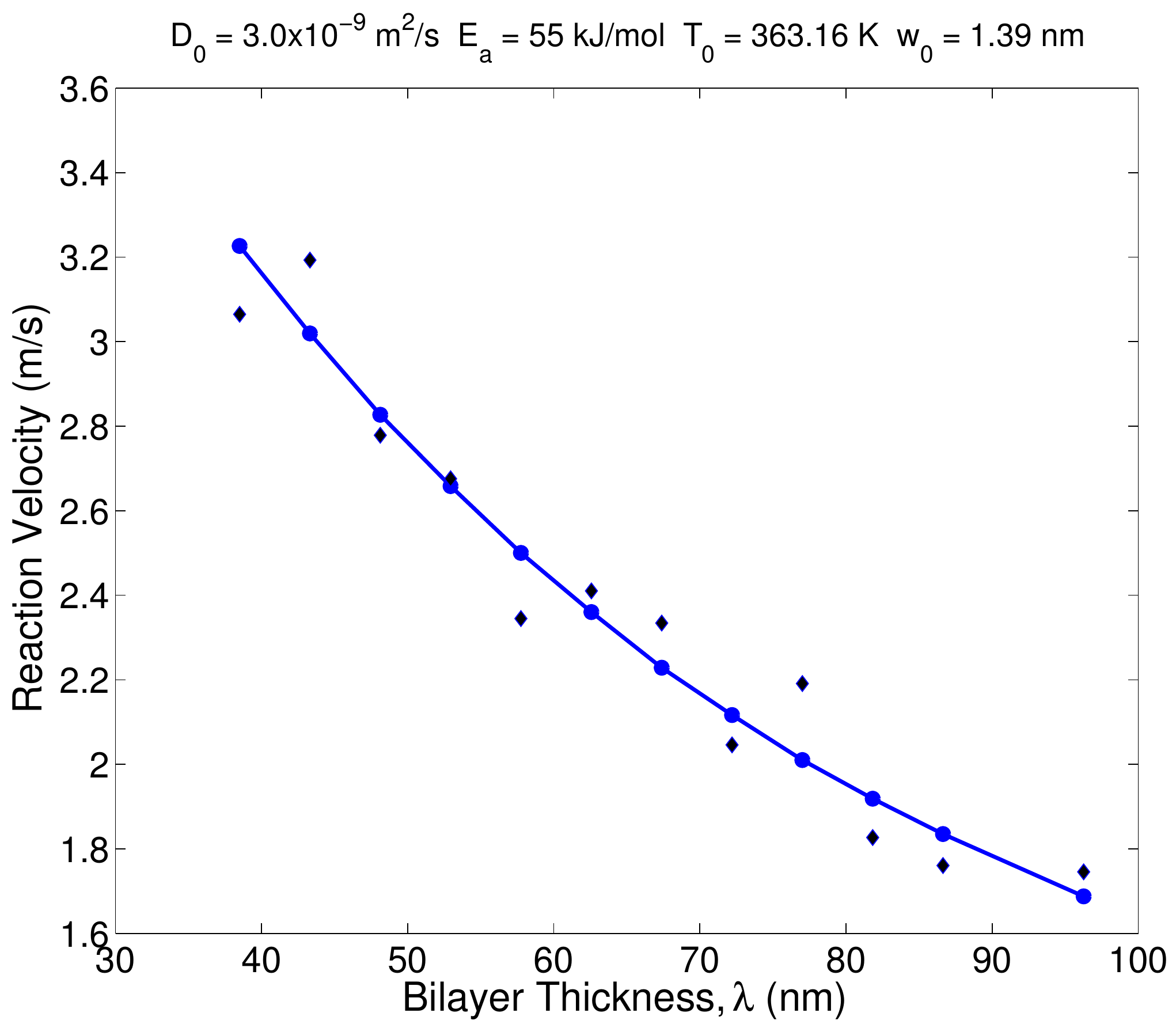}
\caption{Reaction velocity versus bilayer thickness.  The solid curve
  and circular dots corresponds to model 
results generated in the range, 40~nm~$\leq \lambda \leq100$~nm.  The
diamond symbols are 
perturbed values corresponding to a multiplicative Gaussian noise with a standard deviation
of 5\% of the unperturbed model estimate.} 
\label{fig:synth}
\end{center}
\end{figure}

In order to assess the expected information gain associated with
different design scenarios, we generate noisy synthetic data for the
reaction velocity.  The data are generated based on the PC
representation of the model predictions, assuming ``truth'' diffusion
parameter values of $D_0 = 3 \times 10^{-9}$~m$^2$/s and $E_a =
55$~kJ/mol.  The predictions are then perturbed with a Gaussian noise,
using $\alpha = 0.05$ in~\eqref{eqn:disc}.  The resulting synthetic
velocity observations, plotted in Figure~\ref{fig:synth} for a range
of bilayers, are then used as data for the Bayesian calibration step.
As further discussed below, we contrast this $v$-$\lambda$
experimental design with alternatives based on similarly observing the
reaction velocity for a range of initial foil temperatures
($v$-$T_{0}$ design), and a range of premix widths ($v$-$w_{0}$
design).

\begin{figure}[htbp]
 \begin{center}
 \begin{tabular}{cc}
 \hspace{-12mm}
\includegraphics[width=0.52\textwidth]{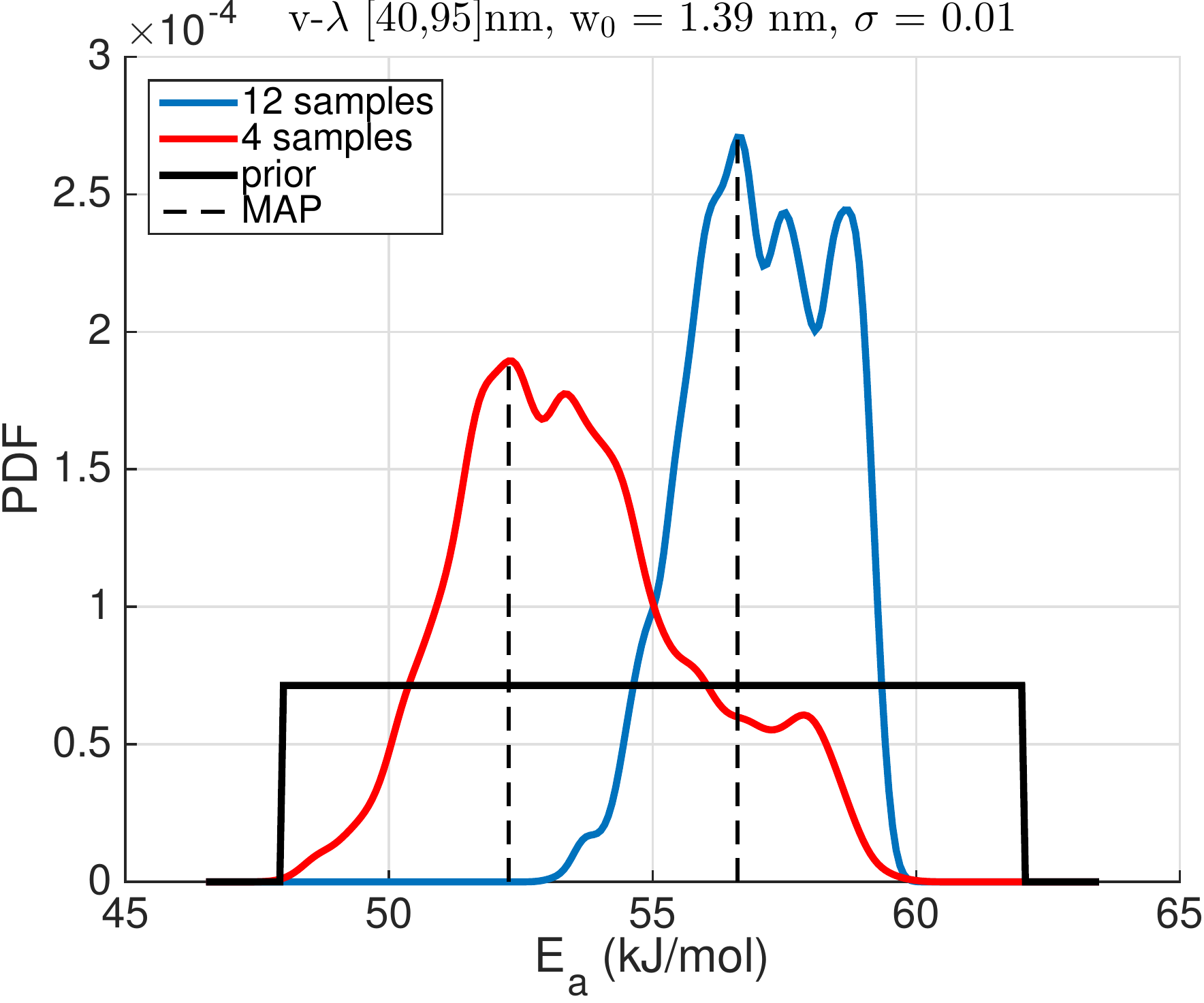}
&
\includegraphics[width=0.52\textwidth]{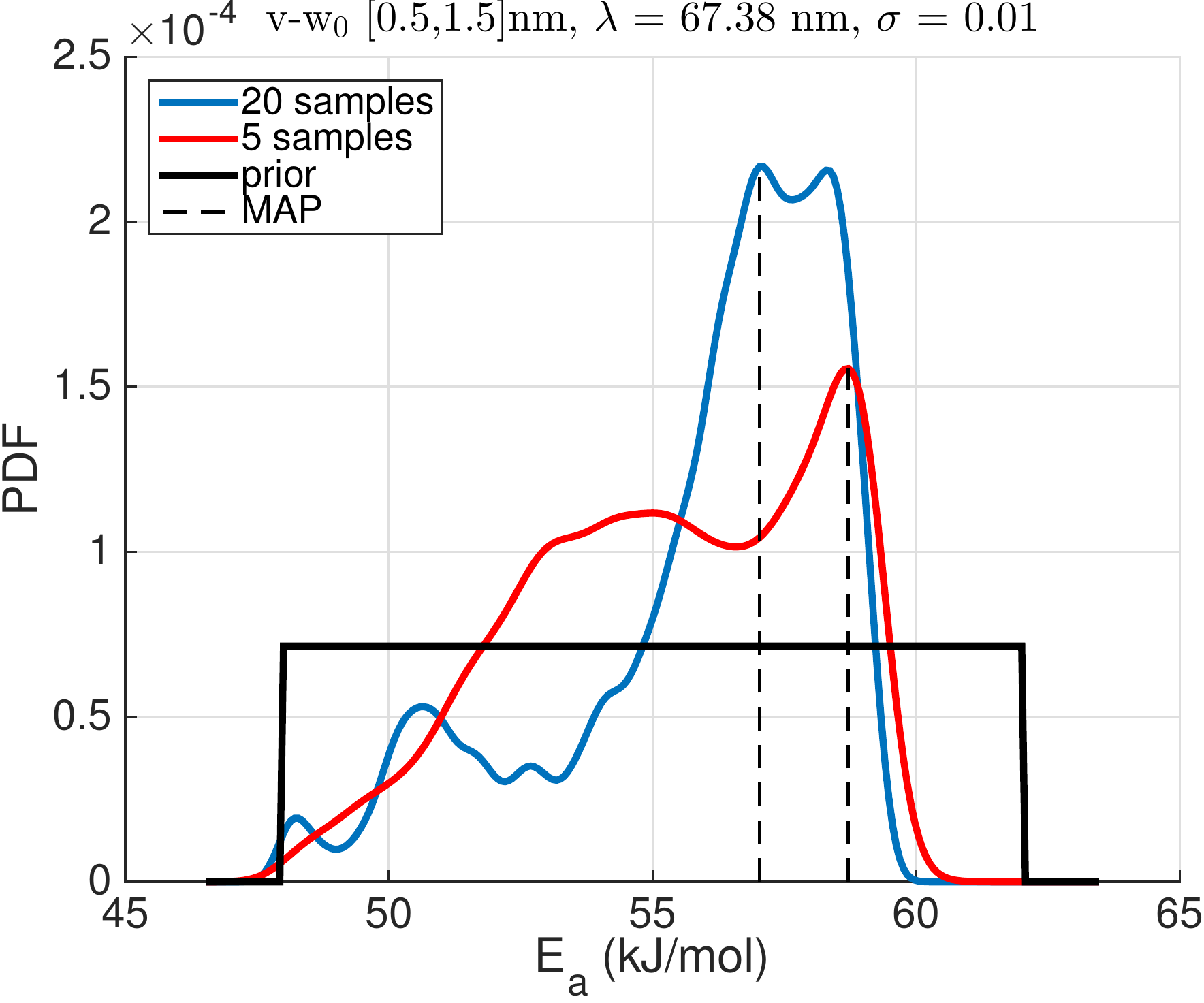}
\\ \vspace{2mm} (a)  & (b) 
\end{tabular}
\caption{Marginal distributions of the posterior of the activation energy.
Results are generated based on (a) $v$-$\lambda$ data pairs, and 
(b) $v$-$w_{0}$ data pairs.  Curves are generated for 
different size data, as indicated.  The uniform
priors and MAP estimates are also plotted.}
\label{fig:samples}
\end{center}
\end{figure}

Figure~\ref{fig:samples} shows posterior marginal distributions of the
activation energy for $v$-$\lambda$ and $v$-$w_0$ designs.  Plotted
are curves generated for different samples sizes, namely 4 and 12 for
the $v$-$\lambda$ design, and 5 and 20 for the $v$-$w_0$ design.  
The results indicate that for both designs, under the present noise
conditions, the larger sample size leads to a significantly sharper
posterior than when the smaller sample size is used.  In addition, the
MAP estimate is closer to the assumed truth value, though evidently a
discrepancy is still observed. The discrepancy arises due to
measurement noise and PC approximation error, and generally would only
vanish in the limit of infinite data and infinite polynomial order in
PC surrogates.

\begin{figure}[htbp]
 \begin{center}
\includegraphics[width=0.70\textwidth]{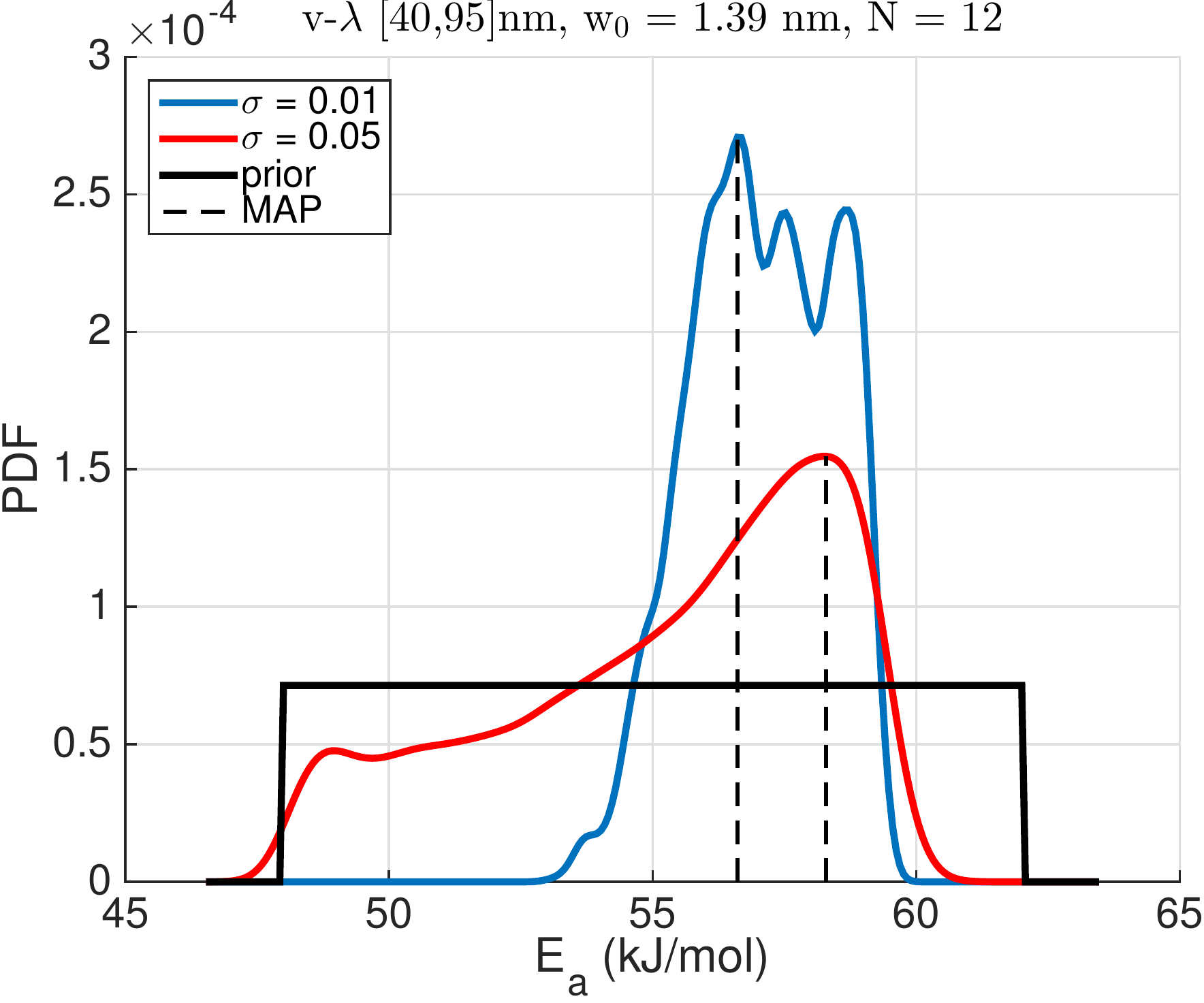}
\caption{Marginal distributions of the posterior of $E_{a}$ for different values of the noise level.} 
\label{fig:noise}
\end{center}
\end{figure}

To assess the impact of observation noise, we generate posterior
marginals of $E_a$ for the $v$-$\lambda$ design and different values
of $\sigma$.  Figure~\ref{fig:noise} contrasts posterior marginals
obtained with $\alpha = 0.05$ and 0.01.  
The distribution
corresponding to a lower noise is observed to be much sharper.  In
other words, we obtain much tighter bounds on the estimates of $E_{a}$
when measurement uncertainty is reduced.  Moreover, it is found that
MAP estimates are closer to the truth value of 55 kJ/mol when sample
size is increased and observation noise is reduced.  Note that in the
results presented above and later in this section, we show posterior
marginals of $E_a$ only, as similar trends exist for $D_{0}$ and so we
omit them for brevity.

\begin{figure}[htbp]
 \begin{center}
\includegraphics[width=0.65\textwidth]{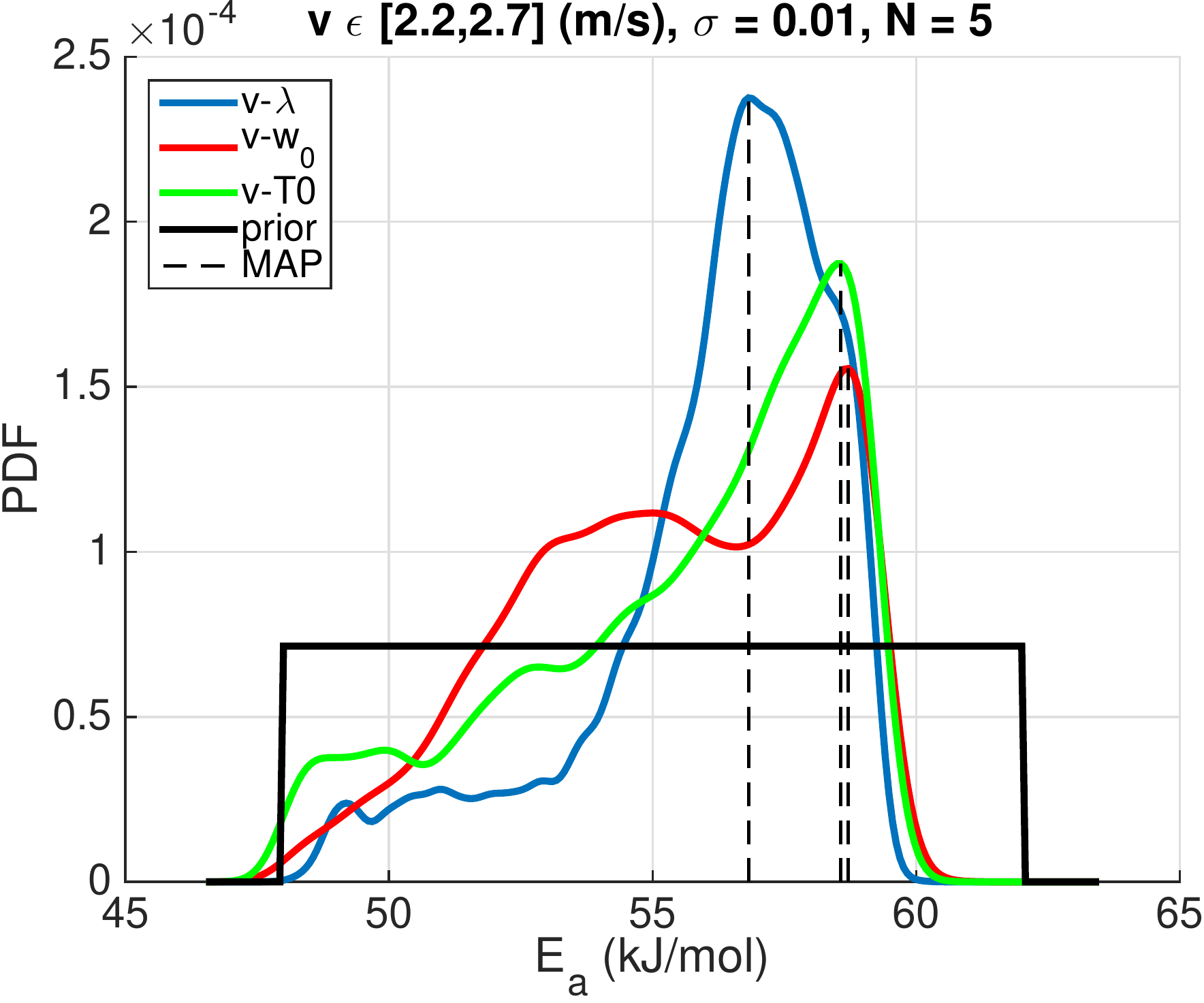}
\caption{Marginal distributions of the posterior of $E_{a}$ based on $v$-$\lambda$,
$v$-$w_{0}$ and $v$-$T_{0}$ data pairs.  The uniform prior and MAP values are
also plotted.} 
\label{fig:choice}
\end{center}
\end{figure}

We now contrast posterior marginals for three different scenarios,
namely based on $v$-$\lambda$, $v$-$w_{0}$ and $v$-$T_{0}$ designs.
In all cases, the sample size ($N$ = 5) and observation noise ($\alpha =
0.01$) are held fixed.  Moreover, the range of values for $\lambda$,
$T_{0}$ and $w_{0}$ are selected such that the reaction velocity
varies in the approximate range 2.2--2.7~m/s.  Figure~\ref{fig:choice}
depicts the marginal posterior distributions of $E_{a}$ for all three
designs.  
The results show that for the case of $v$-$\lambda$
distribution, the information gain is highest followed by $v$-$T_{0}$
and $v$-$w_{0}$. Moreover, the MAP estimate for the $v$-$\lambda$ case
is found to be closest to the assumed truth value, 55 kJ/mol.  Thus,
it is seen that under the present constraints, performing reaction
velocity measurements by varying the bilayer would be more informative
than experimental designs based on varying the initial temperature or
premix width.  As noted earlier, however, varying the bilayer would
generally require a higher effort in synthesizing and characterizing
the requisite number of samples.

As a possible cost-effective alternative to performing experiments at
different bilayer thicknesses, we consider the possibility of
performing repeated experiments at two disparate values of the bilayer
thickness.  This is motivated by recent experiences of optimal
experimental design of shock tube experiments~\cite{Bisetti:2016},
which aimed at maximizing the information gain resulting from a fixed
number of experiments.  The results in~\cite{Bisetti:2016} showed that
an attractive strategy is based on repeated experiments at high and
low values of temperature, namely because these designs are effective
in isolating the impact of observation noise and consequently in
estimating the rate parameters.  In the context of reactive
multilayers, such a strategy would offer the advantage of minimizing
the number of fabrication runs.

\begin{figure}[htbp]
 \begin{center}
 \begin{tabular}{cc}
\includegraphics[width=0.5\textwidth]{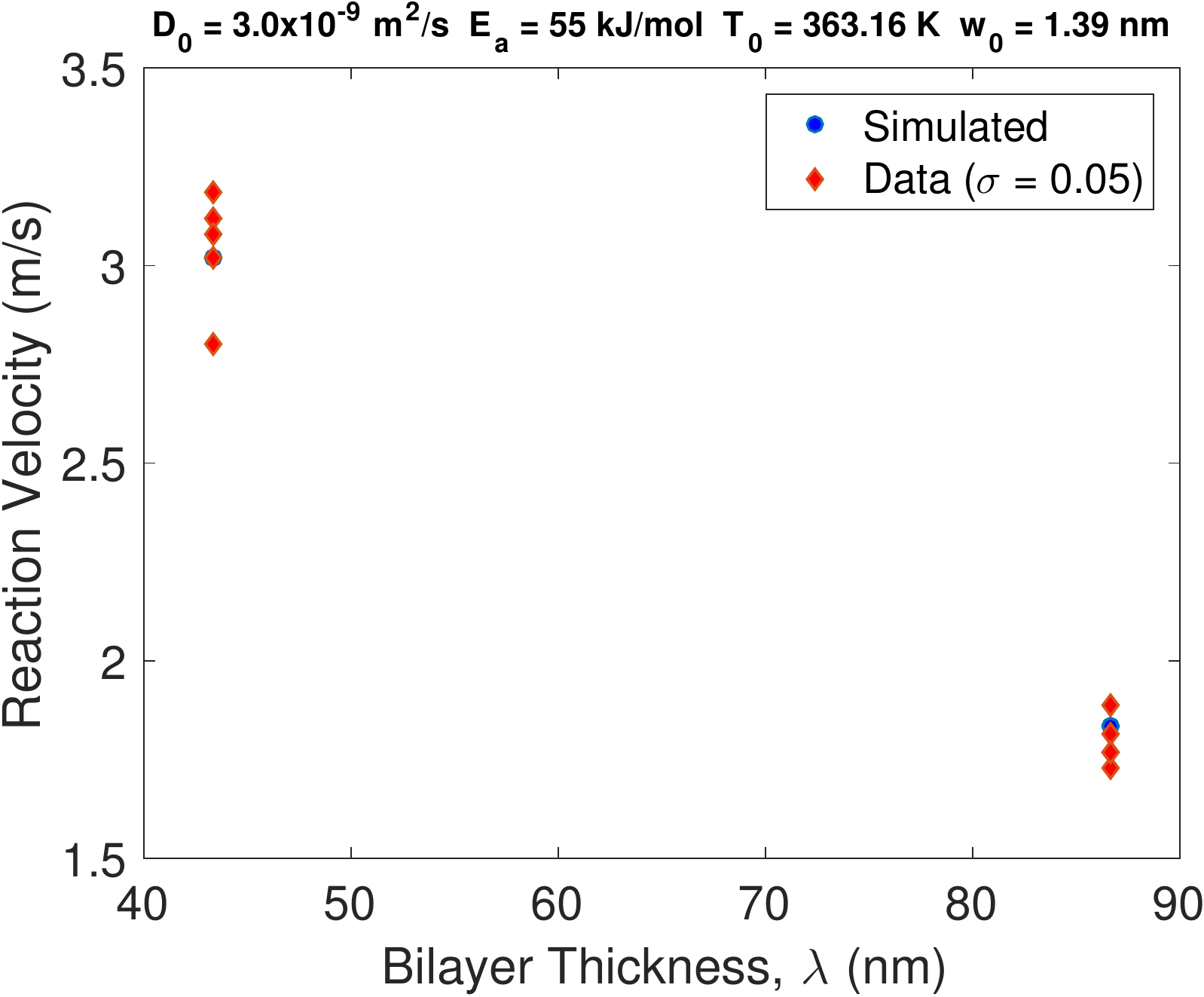}
&
\includegraphics[width=0.5\textwidth]{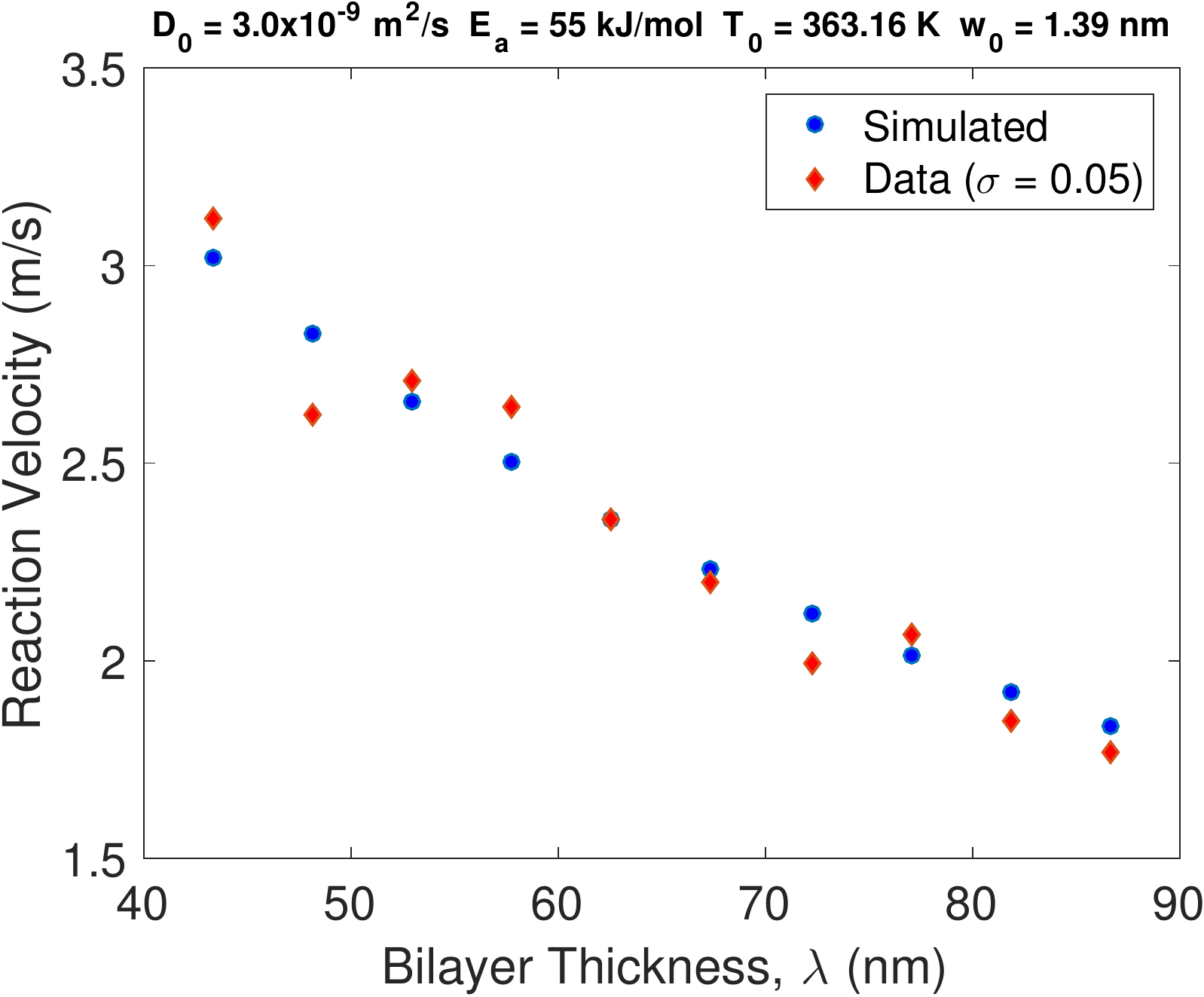}
\\ \vspace{1mm} (a) & (b) 
\end{tabular}
\includegraphics[width=0.5\textwidth]{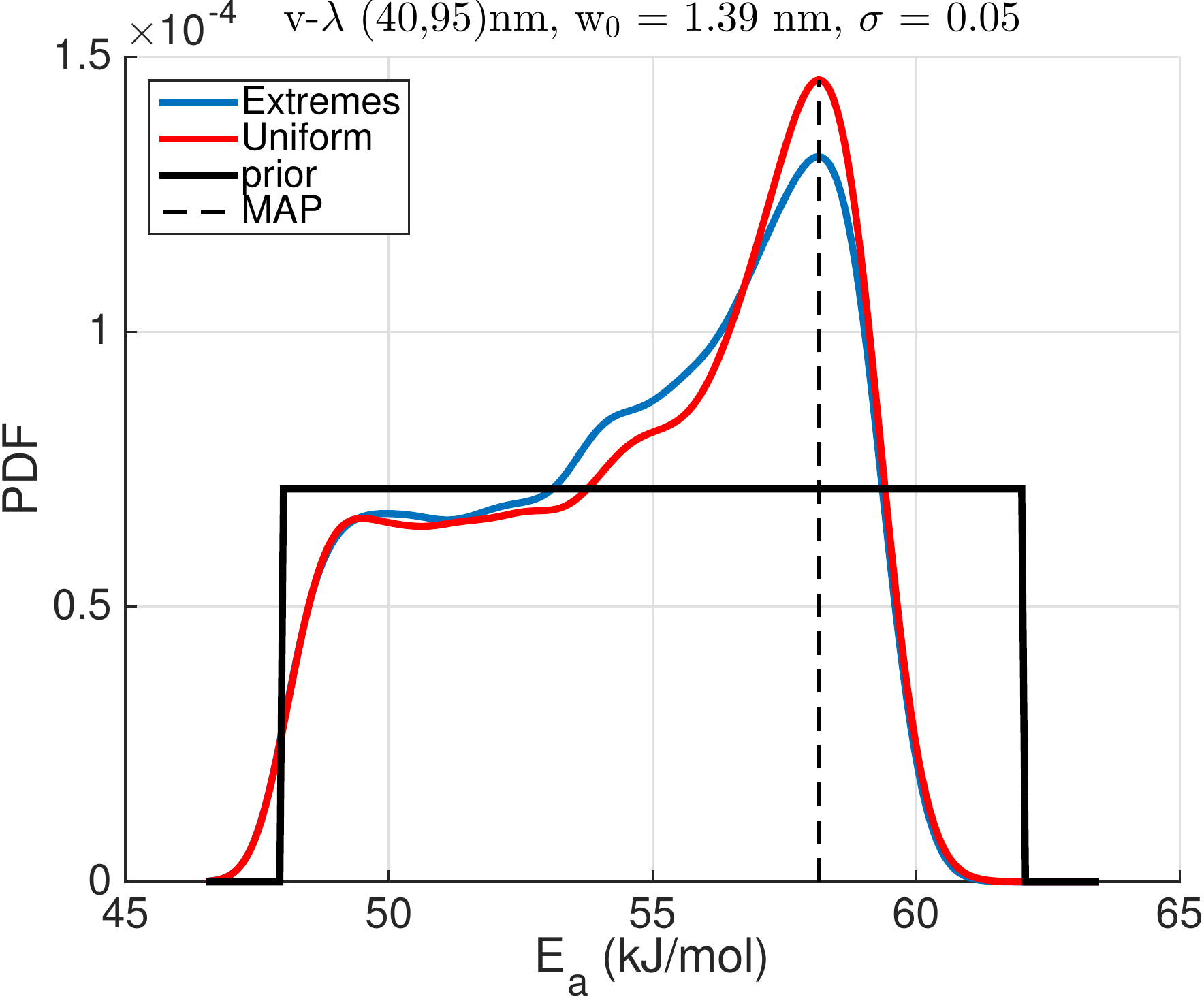}
\\ \vspace{1mm}  (c) 
\caption{(a) Simulated and perturbed reaction velocities for 
$\lambda = 43.32$~nm and $\lambda = 86.64$~nm.  (b) Simulated and 
perturbed reaction velocities for a range of bilayers uniformly distributed
in the range, 43.32~nm~$\leq \lambda \leq 86.64$~nm.
(c)  Marginal distributions of the posterior for $E_{a}$ based on the 
synthetic perturbed data pairs shown in (a) and (b).}
\label{fig:exuni}
\end{center}
\end{figure}

We explore the suitability of this approach by generating multiple
realizations of the synthetic velocity data at low and high values of
the bilayer, and contrast this design to one in which single
observations are made at distinct values of the bilayer that are
uniformly distributed in the range bounded by the low and high bilayer
values.  The synthetic velocity data thus generated are illustrated in
Figures~\ref{fig:exuni}(a) and (b) respectively.  Note that in both
cases the sample size (= 10) is held fixed.  Specifically, in the
repeated experiment design, 5 observations are collected at each of
the high and low values of $\lambda$.
Shown in Figure~\ref{fig:exuni}(c) are the posterior marginal
distributions of the activation energy for the two designs shown in
Figures~\ref{fig:exuni}(a) and (b).  The results indicate that
posterior distributions for both cases are very close.  The
corresponding MAP estimates are also in agreement.  This indicates
that both strategies provide effective means of calibrating the
diffusion parameters, with close estimates of the information gain.
However, as noted above, in the present exercise the repeated
observation design would be more cost effective than the uniformly
distributed bilayer design, because two distinct fabrication runs
would be required in the former whereas ten would be needed in the
latter.

\section{Conclusions}
\label{sec:conc}

We presented a Bayesian approach for systematically designing optimal
reactive multilayer experiments. We focused on Zr-Al multilayers with
a 1:1 molar ratio, and considered both homogeneous ignition and
self-propagation experiments for the goal of learning about the
uncertain Arrhenius parameters that describe the behavior of the
atomic intermixing rates.  For the homogeneous ignition experiment,
the design variables are heating rate and pulse duration, and the
observable is the characteristic time of the transient temperature
profile during the ignition.  For the self-propagation experiment, the
design variables are initial foil temperature and premix width, and
the observable is the front velocity. We also explore the design
effects of bilayer thickness.

The optimal experimental design methodology employs an objective
function (expected utility function) based on the Kullback-Leibler
divergence between posterior and priors, reflecting the expected
information gain on the uncertain parameters from the noisy
measurements at a given design condition; the goal is then to find the
design that maximizes this quantity. The expected utility is estimated
numerically using a doubly-nested Monte Carlo formula. Its evaluation
involves potentially thousands, millions, or even more solves of the
physical model (ODE system), rendering it impractical. We thus
construct a computationally-cheaper surrogate model based on
polynomial chaos (PC) expansions, that capture the dependence of
observables on both design and uncertain parameters.  An adaptive
pseudo-spectral projection technique is used to efficiently build
accurate PC surrogates.  The PC surrogates not only enable tractable
optimal experimental design procedure, but also provide global
sensitivity information and accelerate post-experiment Bayesian
inference/calibration.

For the homogeneous ignition experiment, the design procedure reveals
that the expected utility peaks as the heating rate and pulse duration
decrease.  These predictions are then verified using a Bayesian
calibration exercise, based on synthetic noisy data generated from the
PC surrogate assuming truth values of the activation energy and the
pre-exponent.  The inference is performed at optimal and non-optimal
designs,
and sharper posterior distributions are indeed obtained at the optimal
design conditions.

For the self-propagation experiment, computations indicate that the
expected information gain from experimental reaction velocity
measurements increases as the sample size of the data increases and the
measurement uncertainty decreases.  The analysis also shows that
reaction velocity measurements are more informative when conducted by
varying the bilayer thicknesses as opposed to varying the initial foil
temperature or the initial premix width. Additionally, the two design
scenarios for $\lambda$ (uniformly discretized across a region versus
repeated at two thicknesses) under fixed values of the data size and
measurement uncertainty level and for the same assumed truth
parameters, led to posterior distributions that are in close
agreement.  These results suggest that the repeated measurement approach
can provide a cost-effective alternative to consideration of a large
number of bilayers, as this generally entails a substantial synthesis
cost.

In closing, we note that this study involves a number of assumptions,
such as the lack of model error quantification and the use of an
additive Gaussian noise structure.  While these assumptions are
reasonable for our present purpose, more advanced techniques on these
fronts are certainly of great interest and value, and would be natural
additions to the current framework. Finally, development of the
optimal experimental design formulation and algorithms for multiple
experiments, especially in a sequential manner that makes use of
information feedback, can lead to even greater gains~\cite{Huan:2016}.


\section*{Acknowledgment}

This work was supported by the Defense Threat Reduction Agency, Basic Research Award
\# HDTRA1-11-1-0063, to Johns Hopkins University, and by the US Department of Energy (DOE), 
Office of Science, Office of Advanced Scientific Computing Research, under Award Number
DE-SC0008789. The authors are grateful to Dr. Justin Winokur for providing the aPSP codes 
used to construct model surrogates.


\newpage
\bibliographystyle{unsrt}

\end{document}